\documentclass[11pt,a4article]{article}

\usepackage{jheppub}
\usepackage[T1]{fontenc} 
\usepackage{epsfig,epsf}
\usepackage{amsmath}
\usepackage{amsthm}
\usepackage{amsfonts}
\usepackage{amssymb}
\usepackage{caption}
\usepackage{dsfont}
\usepackage{mathrsfs}
\usepackage{epstopdf}
\usepackage{multirow}

\usepackage{comment}

\usepackage{rotating}

\usepackage{marvosym}

\usepackage{soul}
\usepackage{xcolor}

\usepackage{caption}
\usepackage{subcaption}
\usepackage{array}   
\newcolumntype{C}{>{$}c<{$}}

\usepackage{slashed}

\def\II{\hbox{{1}\kern-.25em\hbox{l}}}

\DeclareMathOperator{\tr}{tr}

\def\II{\hbox{{1}\kern-.25em\hbox{l}}}
\newcommand{\HL}[1]{\textcolor{red}{#1}}

\title{
{\textsc
Anomalous dimensions at small spins}
}

\author[a, b]{A. N. Manashov,}
\author[a]{ S. Moch}
\author[a]{and L. A. Shumilov}

\affiliation[a]{
   II. Institut f\"ur Theoretische Physik, Universit\"at Hamburg,
   D-22761 Hamburg, Germany}
   
\affiliation[b]{Institut f\"ur Theoretische Physik, Universit\"at Regensburg, D-93040 Regensburg, Germany}

\emailAdd{alexander.manashov@desy.de}

\emailAdd{sven-olaf.moch@desy.de}

\emailAdd{leonid.shumilov@desy.de}

\abstract{
In perturbation theory, the anomalous dimensions of twist-two operators have poles 
at negative or small positive integer values of spin and therefore must be resummed at these points.
It was observed earlier that a certain quadratic combination of the anomalous dimensions remains finite at
the right-most singularities, providing an efficient tool for resummation.
In this paper, we analyze the small-spin behavior of the  anomalous dimensions for all types of
twist-two operators in the $O(N)$-symmetric $\varphi^4$ model at the four-loop level,
in the complex $\varphi^3$ model at the three-loop level, and the Gross-Neveu-Yukawa model at the two-loop level. We find that the behavior of the anomalous dimensions at singular points is consistent with theoretical expectations, and we present
expressions for the resummed anomalous dimensions.}

\preprint{DESY-25-077}

\begin{document}
\maketitle

\section{Introduction}

The scale dependence of parton distribution functions relevant to the theoretical description of deep inelastic scattering processes
in quantum chromodynamics (QCD) is governed - via the
operator product expansion - by the anomalous dimensions of twist-two operators.
At present, the latter are known with three-loop accuracy in perturbative QCD~\cite{Moch:2004pa,Vogt:2004mw,Vermaseren:2005qc}
and only partial results are available at four loops~\cite{Moch:2017uml,Gehrmann:2023iah,Gehrmann:2023cqm,Falcioni:2023tzp,Kniehl:2025ttz}.
Higher order calculations remain extremely challenging and, despite significant advances in computational methods,
they can only provide anomalous dimensions for the operators with low values of spin, $s \lesssim 20$, see~\cite{Falcioni:2024qpd} and references therein.
This data is usually insufficient to restore the full analytic $s$ dependence in terms of harmonic sums~\cite{Vermaseren:1998uu} and rational functions of $s$.
Thus, any information about the analytical properties of anomalous dimensions $\gamma(s)$ as functions of spin can be very helpful, since it gives additional constraints. 
For example, the large-$s$ behavior of the anomalous dimensions is determined by the so-called cusp anomalous dimension~\cite{Polyakov:1980ca, Korchemsky:1987wg}
currently known to four-loop accuracy~\cite{Henn:2019swt,vonManteuffel:2020vjv}.
In addition, symmetries such as the reciprocity relation
(RR)~\cite{Dokshitzer:2005bf,Dokshitzer:2006nm,Basso:2006nk} relate the
behavior of the anomalous dimensions $\gamma(s)$ to their values at complementary arguments.
Specifically, the RR imposes constraints on the analytic structure of $\gamma(s)$ under the transformation $s \to -s-1$, reflecting crossing invariance.
As a result, the RR property significantly reduces the number of possible harmonic sums  which can appear in the anomalous dimensions.

By Carlson's theorem, anomalous dimensions continue to an analytic function in the right complex half-plane. In perturbation theory
$\gamma(s)=\sum_k a^k \gamma_k(s)$, where $a$ is a coupling constant and $\gamma_k(s)$ are meromorphic functions with the poles at integer
values of spin. As a rule, the series of poles is bounded from the right with the right-most pole  at $s = s_0$, where $s_0$ is close to $0$ (in known examples $s_0
= -1, 0, 1$ ). At the point $s = s_0$ we expect the behavior $\gamma_k(s) \sim 1/(s - s_0)^{ak + b}$ with $a,b$ being
integers which depend on a model. The perturbative expansion for $\gamma(s)$ obviously fails whenever $a/(s - s_0)\sim 1$. Hence, in order
to determine the correct analytic structure of $\gamma(s)$ near $s\sim s_0$ the perturbative series has to be resummed. In QCD the first
pole in the singlet sector occurs at $s=1$ and corresponds to the BFKL pomeron~\cite{Kuraev:1977fs,Balitsky:1978ic,Jaroszewicz:1982gr,Lipatov:1985uk}.
  
In this work we are interested in the right-most singularity of anomalous
dimensions of leading twist spin-$s$ operators appearing in scalar theories in
$d=4$ and $6$ dimensions.
These singularities share a common feature with the right-most poles of flavor nonsinglet anomalous dimensions in QCD, 
where the small-$s$ behavior is deduced from the double-logarithmic asymptotics of scattering amplitudes~\cite{Kirschner:1982qf, Kirschner:1982xw, Kirschner:1983di}\footnote{The resummation of leading singular contributions in $\varphi^3$ model was done in~\cite{Blumlein:1998mg}.}.
The leading singular contributions to the anomalous dimensions at $s = 0$ were obtained in~\cite{Ermolaev:1995fx, Blumlein:1995jp}.
Several approaches have been developed to resum sub-leading logarithmic contributions~\cite{Velizhanin:2011pb,Vogt:2012gb,Velizhanin:2014dia,Davies:2022ofz}.
In particular, the generalized double-logarithmic equation proposed by Velizhanin~\cite{Velizhanin:2011pb,Velizhanin:2014dia} states that the combination
\begin{align}
\label{DLE}
\delta m^2(s)\equiv 2 \left(s+\bar\beta(a)+\frac12\gamma(s)\right)\gamma(s)
\end{align}
is regular at $s =0$.
Here $\bar\beta(a)=-\beta(a)/2a$ abbreviates the QCD $\beta$-function.
In the planar QCD this expression remains finite at $s\to0$ up to four loops~\cite{Moch:2017uml} 
and in $\mathcal{N}=4$ SYM theory up to seven loops~\cite{Marboe:2016igj}, 
while the non-planar corrections  possess $1/s$ poles  starting from three and four loops, respectively~\cite{Velizhanin:2014dia, Kniehl:2021ysp,Davies:2022ofz,Velizhanin:2022seo}. 
The leading and sub-leading poles, $s^{1-2\ell}$ and $s^{2-2\ell}$ at $\ell$-loops, in the non-planar contributions cancel out.

Recently, some progress has been made in understanding the structure of anomalous dimensions in conformal field theories
(CFT)~\cite{Korchemsky:2015cyx, Caron-Huot:2017vep,Kravchuk:2018htv,Caron-Huot:2022eqs, Li:2025knf, Chang:2025zib}. 
In CFTs the anomalous dimensions (critical dimensions) of operators play a role of observables and do
not depend on the renormalization scheme as in the usual QFT. 
In the $d=4-2\epsilon$ dimensional theory the critical point is determined by the equation $\bar\beta(a_*)=-\epsilon$. 
Thus, the change $\gamma(s,a) \to \gamma(s,a_*)$ in Eq.~\eqref{DLE} provides a statement on CFT observables.
It has been argued in Ref.~\cite{Caron-Huot:2022eqs} that singularities in the anomalous dimensions arise at the points of intersection of two or
more ''Regge'' trajectories.
Specifically,  the trajectory $\Delta(s)$ (or the anomalous dimension $\gamma(s)$) stops to be an analytic function of spin in the vicinity
of an intersection point and acquires square root singularities\footnote{This is similar to the situation of crossing of energy level in ordinary  quantum mechanics.},
which however cancel in the product $ \Delta(s)\widetilde \Delta(s)= \Delta(s)(d-\Delta(s))$. 
It means that $\Delta(s)$ and $\widetilde \Delta(s)$ are different branches of the same analytic function.
The relation in Eq.~\eqref{DLE} is equivalent to the statement that the product $\Delta(s)\widetilde \Delta(s)$ is analytic in the vicinity of the intersection point ($s\sim 0$).
It has been checked that this is indeed true for the twist-two operators in the scalar $\varphi^4$ in $d=4-2\epsilon$ dimensions where the
anomalous dimensions are known with four loop accuracy~\cite{Derkachov:1997pf}, and for the zero- and one-magnon operators in the conformal
fishnet theory in all orders~\cite{Gromov:2018hut}.
For more details see Ref.~\cite{Caron-Huot:2022eqs}. 

It is remarkable that the expression~\eqref{DLE}  arises in another context. 
It was  conjectured by Klebanov and Polyakov~\cite{Klebanov:2002ja} that the critical $O(N)$ vector $\varphi^4$ model in $d=3$ is dual to the higher-spin theory on AdS$_4$,
see also~\cite{Leigh:2003gk,Sezgin:2003pt}.
According to \cite{Girardello:2002pp,Ruhl:2004cf, Manvelyan:2008ks,Skvortsov:2015pea} masses of
the higher-spin fields  on the AdS side are related to the scaling dimensions of twist-two scalar operators ($=$ higher-spin currents) as follows
\begin{align}
m^2(s) = \Delta(s)\big(\Delta(s) -d\big)-s,
\end{align}
where masses are  measured in the units of the cosmological constant. 
Writing $\Delta(s) = d - 2 + s +
\gamma(s)$ and $m^2(s) = (d - 2 + s)(s - 2) - s +\delta m^2(s)$
one  gets
\begin{align}\label{deltam2}
\delta m^2(s) =2\left(s + d/2 - 2 + \frac12\gamma(s)\right)\gamma(s)\,,
\end{align}
which is essentially the same as the expression in~\eqref{DLE}. The critical exponents can be calculated as series in $\epsilon$ in $d=4-2\epsilon$ dimensional theory~\cite{Wilson:1973jj}.
Assuming that the radiative corrections to the mass of the scalar field  ($s=0$) in the dual model are finite one concludes that the r.h.s.
of Eq.~\eqref{deltam2} has to be regular at $s\to 0$. %
Therefore, in the context of  AdS/CFT correspondence one would expect pole cancellation in Eq.~\eqref{deltam2}.

So there are various arguments leading to Eq.~\eqref{deltam2}. The DLE technique~\cite{Kirschner:1982qf,Kirschner:1982xw,Kirschner:1983di}
predicts cancellation of  leading singularities in~\eqref{DLE} in all orders and this is true  in all known examples. The CFT
approach~\cite{Caron-Huot:2022eqs} predicts cancellation of pole singularities. However, there are examples, in QCD and $N=4$ SYM, where taking into account only two trajectories (leading twist and its shadow) is not enough and combination~\eqref{DLE} remains singular. It would be highly desirable, if only for practical purposes, to understand the origin of this difference and
the role  CFT/AdS correspondence plays in the pole cancellation.

In this paper, we analyze the relation~\eqref{deltam2} in several models: the $O(N)$-symmetric vector $\varphi^4$ model, the complex
$\varphi^3$ and Gross-Neveu-Yukawa (GNY) models.
We have calculated the anomalous dimensions for different types of leading twist operators and study
pole cancellation in the corresponding combination of anomalous dimensions. As a result, we provide expressions for resummed anomalous dimensions which remain regular around the right-most singularity points. We also verify that the value at the points coincides with the anomalous dimension of the corresponding local operator if the latter exists.

The paper is organized as follows:
In Sect.~\ref{sect:phi4} we present four loop anomalous dimensions for the scalar, symmetric and anti-symmetric twist-two operators in the $O(N)$-symmetric $\varphi^4$ model and verify that the mass correction~\eqref{deltam2} are free of $1/s$ poles in all cases. 
In Sect.~\ref{sect:phi3} we carry out the same analysis in the complex $\varphi^3$ model in $d=6-2\epsilon$ dimensions in the three-loop approximation. 
We again find that the mass corrections for odd-spin operators are finite at spin minus one. 
In Sect.~\ref{sect:GNY} we analyze GNY model in $d = 4 -2\epsilon$ at two-loop level. 
We calculate anomalous dimensions of all twist-two operators and find the regularity of the mass corrections~\eqref{deltam2} for the multiplicatively renormalazible even-spin operators.
We also discuss the critically equivalent Gross-Neveu model in $1/N$ expansion framework.
We find the regularity of mass corrections~\eqref{deltam2} for the even-spin operators and discuss its connection to the intersection of Regge trajectories at the points $s = \pm(2 - \mu)$ and $s = 0$, where $\mu = d/2$.
The last Section contains our conclusions. 
The Appendix~\ref{app:thre-loop-res} contains explicit three-loop expressions for the anomalous dimensions in the complex $\varphi^3$.

\section{$O(N)$-symmetric $\varphi^4$ model}\label{sect:phi4}
The $O(N)$-symmetric $\varphi^4$ model played an important role in the development of the theory of critical phenomena and provides a
basic example of a nontrivial CFT. It represents a natural  testing ground for new approaches and methods,
see e.g.~\cite{Vasiliev:1982dc,Rattazzi:2008pe,Alday:2015eya,Panzer:2016snt,Badel:2019oxl}. See also Ref.~\cite{Henriksson:2022rnm} for the recent review. 
The model describes dynamics of an  $N$ component scalar field $\varphi^a$, $a=1,\ldots,N$, with an $O(N)$ invariant
quartic interaction,
\begin{align}\label{phi4action}
    S_{R} = \int d^dx\left(\dfrac{Z_1}{2}\left(\partial\varphi\right)^2 + Z_3\mu^{2\epsilon}\dfrac{g}{4!}\left(\varphi^2\right)^2\right).
\end{align}
Here  $Z_{1,3}$ are the renormalization constants in the $\overline{\text{MS}}$ scheme and $\mu$ is the renormalization scale. The
renormalization group (RG) functions in this model are known with high-precision, see~\cite{Kazakov:1979ik, ZinnJustin:1989mi,Vasilev:2004yr, Kompaniets:2017yct,Schnetz:2022nsc} for recent developments.
In particular the beta function and the field anomalous dimension, which we will need later, take the form
\begin{align}
\beta(u) &= \mu\dfrac{d u}{d \mu} 
=-2\epsilon u + \dfrac{N + 8}{3}u^2 - \dfrac{3N + 14}{3}u^3
\notag\\
&\quad \notag
+\frac{u^4}{216}\Big(33N^2+922N+2960+96(5N+22)\zeta_3\Big) + O(u^5)\equiv -2u(\epsilon +\bar\beta(u)),
\notag\\
\gamma_\varphi(u) &=\mu\dfrac{d\ln Z_1 }{d \ln\mu^2} = (N+2)\Bigg(\frac{u^2}{36}-\frac{u^3(N+8)}{432}
+\frac{5u^4}{5184}(-N^2+18N+100)\Bigg)+O(u^5),
\end{align}
where $u=g/(4\pi)^2$.

We are interested in the anomalous dimensions of twist-two operators 
\begin{equation}
\label{twist-2-def}
\mathcal O_s^{ab} =   \mathcal{O}_{\mu_1\ldots\mu_s}^{ab}(x) = \varphi^a(x)\partial_{\mu_1}\ldots\partial_{\mu_s}\varphi^b(x) - \text{traces}.
\end{equation}
It is convenient to single out irreducible components relative to the $O(N)$ group. Namely, we define the scalar, $\mathcal O_s=\sum_a
\mathcal O_s^{aa}$,  symmetric-traceless, $\mathcal O^+_s= \frac12(\mathcal O_s^{ab}+\mathcal O_s^{ba})- N^{-1}\delta^{ab} \mathcal O_s$, and
anti-symmetric, $\mathcal O^-_s= \frac12(\mathcal O_s^{ab}-\mathcal O_s^{ba})$, operators which do not mix under renormalization,
\begin{align}
\label{z-matrix-phi4}
    \left[\mathcal{O}_s^{\alpha}(x)\right]_{\overline{\text{MS}}} = Z^{\alpha}_s(u,\epsilon)\mathcal{O}_s^{\alpha}(x).
\end{align}
Here $\alpha$ denotes the type of operator (symmetric, anti-symmetric  or scalar) and $Z^{\alpha}_s=1 +\sum_{k=1}^\infty \epsilon^{-k}
{z^\alpha_k(s,u)}$ are the renormalization constants. The anomalous dimensions are given by the following expression
\begin{align}
\gamma^\alpha(s) &= 2\gamma_\varphi - \mu \frac d{d\mu}  \ln Z^{\alpha}_s(u,\epsilon),
\end{align}
where $\gamma_\varphi=\frac12 \mu \partial_\mu \log Z_1$ is the anomalous dimension of the field $\varphi$ (wave function renormalization).
For $N=1$ theory the anomalous dimensions, $\gamma(s)$, were calculated in the four loop approximation in Ref.~\cite{Derkachov:1997pf}. 
This result was generalized for the $O(N)$ model in Ref.~\cite{Manashov:2017xtt} by recovering  the symmetry coefficients for individual graphs 
\allowdisplaybreaks{
\begin{align}\label{u-expansion-sc}
\gamma(s)& =2\gamma_\varphi- u^2\frac{N+2}{9s(s+1)}\Biggl\{
3  + \frac{u}{3}\big(N + 8\big)
\left(2S_1- 5 +\frac32 \frac{2s+1}{s(s+1)}\right)
+\dfrac{u^2}{9}\Biggl(\big(N + 8\big)^2S_2 \notag\\
&\quad + \big(N^2 + 11N + 42\big)S_1^2 + \big(5N + 22\big)\biggl(1 - \dfrac{4}{s(s + 1)}\biggr)S_{-2} + \frac{3(N+2)}{2}\biggl[S_1\!
-\!\frac{11}2 \notag\\
&\quad
\!+3\frac{2s + 1}{s^2 (s+1)^2}+\frac{s-3}{s (s+1)}
\biggr]
\!
+\big(N^2\!+6N\!+20\big)\bigg[
\biggl(\dfrac{1 + 2s}{s(s + 1)} - 4\biggr)S_1
 + \frac{3}{4 s^2 (s+1)^2}\notag\\
&\quad
-\frac{20s + 1}{4s(s+1)} + 2\bigg]
+\big(5N+22\big)\biggl[
S_1\biggl(2\frac{2s + 1}{s(s+1)} - 11\biggr)
+\frac{3}{2 s^2 (s+1)^2}
\notag\\
&\quad
- \frac{32s + 7}{2s(s+1)} + 21
 \biggr]\Biggr)
 \Biggr\} + O(u^5)\,.
\end{align}
Here $S_{k} = S_{k}(s)$ are the harmonic sums~\cite{Vermaseren:1998uu}. 
Doing the same for the symmetric and anti-symmetric operators we obtain the following anomalous dimensions
\begin{align}
\label{u-expansion-st}
\gamma^+(s)& =2\gamma_\varphi-\frac{u^2}{9s(s+1)}\Biggl\{N+6
+\frac{u}6\Biggl( {8\left(N + 4\right)}\left[2S_{1} + \dfrac{2s + 1}{s(s + 1)} - 4\right]
- \big(N^2 + 6N \notag\\
&\quad + 16\big)\biggl[{2} - \frac{2s+1}{s(s + 1)}\biggr]\Biggr)
 +\dfrac{u^2}{9}\Biggl(4\big(N + 4\big)\big(N + 8\big)S_2
+  \big(N^2 + 28N + 84\big)S_1^2 \notag\\
&\quad + \biggl(N^2 + 16N + 44 - 8\frac{5N + 22}{s(s + 1)}\biggr)S_{-2}
+\big(N^3 + 8N^2 + 24N + 40\big)\biggl[\frac{1}{4 s^2 (s+1)^2} \notag\\
&\quad- \frac{4s-1}{4s(s+1)}\biggr]
+\big(N^2 + 16N + 44\big)\bigg[
 \biggl(\frac{5}{3}\frac{2s + 1}{s(s+1)} - \frac{32}{3}\biggr)S_1
+ \frac{5}{6s^2(s+1)^2}
\notag\\
&\quad - \frac{64s + 17}{6s(s+1)} + \frac{52}{3}\bigg]
 +\dfrac{3N^2 + 20N + 44}{3}\biggl[
 \biggl(\dfrac{2s + 1}{s(s + 1)} - 1\biggr)S_1
+ \frac{2}{s^2(s+1)^2} \notag\\
&\quad - 2\frac{8s + 1}{s(s+1)} + 11\biggr]
 + \dfrac{N + 6}{2}\biggl\{\left(N + 2\right)\bigg[S_1
+ \dfrac{s}{s(s + 1)}
 - \dfrac{11}{2}\bigg] + \big(N + 6\big)\biggl[\frac{2s + 1}{s^2 (s+1)^2}\notag\\
&\quad - \frac{1}{s (s+1)}\biggr]
+\dfrac{8(N + 5)}{3}\biggl[
\biggl(\dfrac{2s + 1}{s(s + 1)} - 4\biggr)S_1 + \frac{1}{2 s^2 (s+1)^2} - \frac{8s + 1}{2s (s+1)}  \notag\\
&\quad +2\biggr] \biggr\}\Biggr)
\Biggr\} + O(u^5)\,,
\end{align}
\begin{align}
\label{u-expansion-as}
\gamma^-(s) &= 2\gamma_\varphi -u^2\frac{N+2}{9s(s+1)}\Biggl\{1+u\Biggl(\frac23\left[2S_{1} + \dfrac{2s + 1}{s(s + 1)} - 4\right]
- \frac{N+4}6\left[{2} - \frac{2s+1}{s(s + 1)}\right]
\Biggr) \notag \\ &\quad
+\dfrac{u^2}{9}\Biggl(2\big(N+8\big) S_2 -\big(N - 6\big)S_1^2 - \big(N-2\big)S_{-2}
+\big(N^2 + 6N + 12\big)\biggl(\frac{1}{4 s^2 (s+1)^2}\notag \\ &\quad - \frac{4s-1}{4s(s+1)}\biggr)
+ \dfrac{4\big(6 + N\big)}{3}\biggl[
\biggl(\dfrac{2s + 1}{s(s + 1)} - 7\biggr)S_1
+ \frac{1}{2 s^2 (s+1)^2} - \frac{7s + 2}{s(s+1)}+\frac{25}{2}\biggr]
\notag \\ &\quad
+\dfrac{10 + 3N}{3}\biggl[
\biggl(\frac{2s+1}{s (s+1)} - 1\biggr)S_1 + \frac{2}{s^2 (s+1)^2}-2\frac{8 s+1}{s (s+1)}+11\biggr]
\notag \\ &\quad + \dfrac{14 - N}{3}\biggl[
\biggl(\frac{2 s+1}{s (s+1)}-4\biggr)S_1  + \frac{1}{2s^2(s+1)^2} - \frac{8s + 1}{2s(s+1)} + 2\biggr]
+ \dfrac{N + 2}{2}\biggl[S_1 \notag \\ &\quad + \frac{2 s+1}{s^2 (s+1)^2}+\frac{s-1}{s(s+1)}-\frac{11}{2}\biggr] 
\Biggr)
\Biggr\} + O(u^5)\,.
\end{align}
One can easily check that the anomalous dimensions corresponding to the stress-energy tensor and the conserved $O(N)$ current vanish, 
$\gamma(2)=0$ and $\gamma^{-}(1)=0$. These results agree with the Ref.~\cite{Henriksson:2018myn}, where they were obtained using Lorentzian inversion formula.

We have also checked that the anomalous dimensions $\gamma^\alpha(s)$ have the so-called reciprocity respecting
form~\cite{Dokshitzer:2006nm, Korchemsky:1987wg, Alday:2015eya}.
Namely, the functions $f^\alpha(j)$ defined by the equation
\begin{align}
\gamma^\alpha(s) =f^\alpha\left(j\right),
\end{align}
where $j = s + \bar{\beta}(u) +{\frac12 \gamma(s)}$ is the conformal spin, is "almost" invariant under the $j \to -j - 1$,
for details see Refs.~\cite{Dokshitzer:2006nm, Korchemsky:1987wg, Alday:2015eya}.
For the operators in question we obtain\footnote{Note that the expression~\eqref{DLE} can be written in the form $\delta m_s^2(s) = 2jf(j)$.}
\begin{align}
\label{f-u-expansion-sc}
    f(j) &= 2\gamma_{\varphi} - u^2\dfrac{N + 2}{9j(j + 1)}\Biggl\{3 + \dfrac{2u}{3}\big(N + 8\big)\biggl(S_1 - \dfrac{5}{2}\biggr)
    +\dfrac{u^2}{9}\Biggl(\big(22 + 5N\big)\biggl(1 - \dfrac{4}{j(j + 1)}\biggr)S_{-2}
     \notag \\ &
     \quad + \big(N^2 + 11N + 42\big)S_1^2 - \dfrac{16N^2 + 310N + 1276}{4}S_1 + \big(N + 8\big)^2\zeta_2 - \dfrac{21(N + 2)}{4j(j + 1)}
      \notag \\ &
      \quad+ \dfrac{8N^2 + 435N + 1942}{4}\Biggr)\Biggr\} + O(u^5),
\\
\label{f-u-expansion-st}
    f^{+}(j) &= 2\gamma_{\varphi}  - u^2\dfrac{1}{9j(j + 1)}\Biggl\{N + 6 + \dfrac{u}{3}\Biggl(8(N + 4)S_1 - N^2 - 22N - 80\Biggr)
    \notag \\ &\quad
    +\dfrac{u^2}{9}\Biggl(\left(N^2 + 16N + 44 - 8\dfrac{22 + 5N}{j(j + 1)}\right)S_{-2} + \big(N^2 + 28N + 84\big)S_1^2
    \notag \\ &
    \quad
    -\dfrac{33N^2 + 464N + 1276}{2}S_1 + 4\big(N + 4\big)\big(N + 8\big)\zeta_2 - \dfrac{3N^2 + 32N + 84}{4j(j + 1)}
    \notag \\ &
    \quad + \dfrac{113N^2 + 1432N + 3884}{4}\Biggr)\Biggr\} + O(u^5),
\\
\label{f-u-expansion-as}
    f^{-}(j)& = 2\gamma_{\varphi} - u^2\dfrac{N + 2}{9j(j + 1)}\Biggl\{1 + \dfrac{u}{3}\Biggl(4S_1 - 12 - N\Biggr)
    - \dfrac{u^2}{9}\Biggl((N - 2)S_{-2} + (N - 6)S_1^2
    \notag \\
    &\quad
    +\dfrac{17N + 154}{2}S_1 - 2(N + 8)\zeta_2
     + \dfrac{3\left(N + 2\right)}{4j(j + 1)} - \dfrac{97N + 562}{4}\Biggr)\Biggr\} + O(u^5).
\end{align}
These expressions contain only reciprocity respecting combinations of harmonic sums~\cite{Beccaria:2009vt, Beccaria:2010tb}, 
while all rational terms are expressed through the RR invariant combination $j(j + 1)$.

\subsection{Small $s$ limit}

The anomalous dimensions of all operators are singular in the limit  $s \to 0$. At $\ell$ loops one expects to find contributions which
diverge as $s^{-\ell+1}$ when $s \to 0$. Using the results of the previous Section it is easy  to verify that the combination
\begin{align}\label{m2}
\delta m_{\alpha}^2(s)=2\left(s+\bar\beta(u)+\frac12 \gamma^{\alpha}(s)\right)\gamma^{\alpha}(s)
\end{align}
remains finite in the limit $s\to 0$ up to $O(u^5)$ terms for all operators\footnote{The $N=1$ case was considered in Ref.~\cite{Caron-Huot:2022eqs}.}. Note that in order to obtain the  limit $s \to 0$ in Eqs.~\eqref{u-expansion-sc}, \eqref{u-expansion-st} and~\eqref{u-expansion-as} one needs to analytically continue the harmonic sums to non-integer values of $s$. Techniques for such a continuation are well developed. For a detailed discussion, as well as for the corresponding \texttt{MATHEMATICA} package, see, e.g., Refs.~\cite{Velizhanin:2020avm, Velizhanin:2022ays}.
Moreover, for the scalar and symmetric operators the Regge trajectories, as predicted in~\cite{Caron-Huot:2022eqs}, pass through the point corresponding
to the local operators, $\mathcal O_{s=0}=\varphi^2$, and $\mathcal O^{+}_{s=0}=\varphi^a\varphi^b-N^{-1}\delta^{ab}\varphi^2$. 
The anomalous dimensions, $\gamma_{0}=\gamma_{\varphi^2}$ and  $\gamma_0^+$, are known with high accuracy, see e.g.~\cite{Vasilev:2004yr}
\begin{align}
    \gamma_0 =& u\dfrac{N + 2}{3}\Biggl\{1 - \dfrac{5u}{6} + u^2\dfrac{(37 + 5N)}{12}\Biggr\} + O(u^4), \\
    \gamma_0^+ =& \dfrac{u}{3}\Biggl\{2 - u\dfrac{10 + N}{6} - u^2\dfrac{5N^2 - 84N - 444}{72}\Biggr\} + O(u^4).
\end{align}
Defining
\begin{align}
\label{m2-zero}
\delta m_0^2&\equiv2\left(\bar\beta(u)+\frac12 \gamma_{0}\right)\gamma_{0}
        =2u^2(N + 2)\Biggl\{{-}\dfrac{1}{3}+ u\dfrac{13(N + 8)}{108}
        \notag\\
    &\quad
    - u^2\left(\dfrac{2(22 + 5N)}{27}\zeta_3 + \dfrac{3N^2 + 950N + 4588}{1296}\right)\Biggr\} + O(u^5),
\notag\\
\delta m_0^{2,+}&\equiv2\left(\bar\beta(u)+\frac12 \gamma_{0}^+\right)\gamma_{0}^+
= 2u^2\Biggl\{-\dfrac{N + 6}{9} + u\dfrac{N^2 + 50N + 208}{108}
 \notag \\ &
    \quad-u^2\left(\dfrac{4(22 + 5N)}{27}\zeta_3 - \dfrac{5N^3 - 164N^2 - 3112N - 9176}{1296}\right)\Biggr\} + O(u^5)
\end{align}
we have checked that with four loop accuracy
\begin{align}\label{m2to0}
\lim_{s\to 0}\delta m^2(s) = \delta m_0^2\,, &&\lim_{s\to 0} \delta m_{+}^{2}(s) = \delta m_0^{2,+}\,.
\end{align}
The situation is similar for the anti-symmetric operator, $\mathcal O_s^{-}$, although  there is no local operator for $s=0$.
We have found that in the  limit $s\to 0$  the mass correction remains finite
\begin{align}
\label{m2-odd}
   \lim_{s\to 0} \delta m^2_{-}(s) &= 2u^2(N + 2)\Biggl\{-\dfrac{1}{9}
    + u\dfrac{24 + N}{108}+  u^2\left(\dfrac{5N^2 - 158N - 1148}{1296}\right)\Biggr\} + O(u^5).
\end{align}
As we will see in Sect.~\ref{sect:phi3} the combination in Eq.~\eqref{deltam2} remains singular at $s=0$ if anomalous dimensions are continued from {\it odd} values of $s$.
Therefore, the finiteness of $\delta m_{-}^{2}(s)$ in the first four loops requires an additional explanation. In
Sect.~\ref{sect:largeN} we argue that this fact can be better understood in the framework of the $1/N$ expansion.

\vskip 2mm

Note also that Eq.~\eqref{m2} defines the anomalous dimension $\gamma(s)$ for small $s$ as an analytic function on a
two-sheet Riemann surface, see also \cite{Velizhanin:2014dia,Caron-Huot:2022eqs},
\begin{align}
\label{analyt-cont-def}
\gamma(s)=-s-\bar \beta + \sqrt{(s+\bar\beta)^2+ \delta m^2(s)}.
\end{align}
The position of the branching points is determined by the expression under the square root.
For the $O(N)$ model at the critical point, $\bar\beta=-\epsilon$,
both branching points are of order $\epsilon$ and are located on the positive axis ($\delta m^2 < 0$ as we see in Eqs.~\eqref{m2-zero},~\eqref{m2-odd}).
One branch passes through the anomalous dimensions of the corresponding  twist-two
operators at integer $s$ and another one through the anomalous dimensions of the shadow operators.

 \vskip 3mm

Regularity of the mass correction in Eq.~\eqref{m2} can be used as a constraint on the expansion of anomalous dimensions at $s = 0$.
Knowing anomalous dimensions at $\ell$-th loop order one can find all singular terms except $1/s$  in the next order. 
The simple pole can be restored from the anomalous dimension of the local operator~\eqref{m2-zero}. 
In this way we obtain the five-loop  singular part of the anomalous dimension of the scalar operator (the necessary results for the four-loop beta function and $\gamma_{\varphi^2}$ can be found in~\cite{Vasilev:2004yr})
\begin{align}
    \gamma^{(5)}(s) &= \left(N + 2\right)\Biggl\{-\dfrac{(N + 8)(N^2 + 34N + 100)}{648s^4}
    +  \dfrac{(N + 8)(5N^2 + 236N + 776)}{972s^3}
    \notag \\
    &\quad
    -\Biggl(\dfrac{(14 + N)(22 + 5N)}{81}\zeta_3 + \dfrac{(8 + N)(N^2 + 28N + 88)}{486}\zeta_2
    \notag \\
    &\quad + \dfrac{16N^3 + 2067N^2 + 24270N + 63152}{3888}\Biggr)\dfrac{1}{s^2}
    + \Biggl(\dfrac{40N^2 + 1100N + 3720}{243}\zeta_5
    \notag \\
    &\quad
    + \dfrac{(32 + N)(22 + 5N)}{810}\zeta_2^2 - \dfrac{5N^3- 365N^2- 5278N  -16880}{972}\zeta_3
    \notag \\
    &\quad + \dfrac{(N + 8)(N + 26)(8N + 35)}{972}\zeta_2 + \dfrac{371N^2 + 6286N + 18192}{648}\Biggr)\dfrac{1}{s}\Biggr\} + O(s^0).
\end{align}
In the case $N = 1$ this prediction is consistent with the one obtained in~\cite[Eq. (2.28)]{Caron-Huot:2022eqs}.

The $O(N)$-symmetric  $\varphi^4 $ model can be also analyzed in the
$1/N$ expansion, see e.g.~\cite{ZinnJustin:1989mi,Vasilev:2004yr} for a review. The corresponding anomalous dimensions for the symmetric
and anti-symmetric operators were calculated in Ref.~\cite{Derkachov:1997ch}, and for the scalar operators in Ref.~\cite{Manashov:2017xtt}.
We have checked that our expressions in Eqs.~\eqref{u-expansion-sc},~\eqref{u-expansion-st} and~\eqref{u-expansion-as} agree with these results. 
In the $1/N$ expansion the anomalous dimensions are regular functions at $s=0$ but they acquire a  pole at the point $s = d/2 - 2$.
This pole, however, cancels in the mass corrections, analogous to Eq.~\eqref{deltam2}. The explicit example for the scalar operator in $d=3$ can be found in~\cite[Eq.~(4.2)]{Manashov:2017xtt}.
Therefore $\delta m_{\alpha}^2(s)$ are regular functions for $s \ge 0$ and the corresponding anomalous dimensions are defined
analogous to Eq.~\eqref{analyt-cont-def} around the point $s = d/2 - 2$.

\section{Complex cubic models\label{sect:phi3}}
In this Section, we study the mass corrections~\eqref{deltam2} in two six-dimensional models. 
The first model  describes {the} self-interacting  complex field $\varphi$
\begin{align}\label{varphi3complex}
S_I&=\int d^dx \left(\partial\varphi\partial\bar\varphi +\frac16 g\Big(\varphi^3    + \bar \varphi^3 \Big)\right)
\end{align}
and the second model involves an $N$ component field $\varphi=\{\varphi_1,\ldots, \varphi_N\}$ and {the} scalar field $\sigma$
\begin{align}
\label{varphi31N}
S_{II}&=\int d^dx \left(\partial\varphi\partial\bar\varphi + \partial\sigma \partial\bar\sigma + \frac12 g\Big(\varphi^2\sigma
    +\bar\varphi^2\bar \sigma \Big)\right)\,,
\end{align}
where $\varphi^2=\sum_a \varphi_{a}^2$.
They are special cases of the general cubic models considered in
Refs.~\cite{deAlcantaraBonfim:1980pe,deAlcantaraBonfim:1981sy,Gracey:2020tkk,Fei:2014xta,Fei:2014yja,Borinsky:2021jdb}. 
Model~I is invariant under the discrete symmetry transformations, $\varphi\mapsto e^{\pm i2\pi/3}\varphi$,  that guarantees its multiplicative renormalizability. The
Model~II, in addition to $O(N)$ symmetry is invariant under $U(1)$ rotations $\varphi\mapsto e^{i\alpha}\varphi$, $\sigma\mapsto e^{-2i\alpha}\sigma$. 
For $N=2$ the symmetry  is enhanced to $U(1) \times U(1)\times S_3$ and the action can be written as
\begin{align}
%
S&=\int d^dx
 \left(\sum_{{k}=1}^3\partial\varphi_{{k}}\partial\bar\varphi_{{k}}
 +  g\big(\varphi_1\varphi_2\varphi_3+\text{c.c}\big)\right)\,.
\end{align}
    The renormalization group functions read ( $u=g^2/(4\pi)^3$ )
\begin{align}
\gamma_\varphi(u) &=  \frac u{12}  - \frac{11}{3}\left(\frac u {12} \right)^2
		+\frac{u^3}{24}\left(\frac{5357}{2592}-\zeta_3\right)
		+ O(u^4)\,,
\notag\\[2mm]
\beta(u)&= -2\epsilon u+ \frac12 u^2 -\frac{83}{72}u^3+O(u^4)
\end{align}
for model~I and
\begin{align}
\gamma_\sigma(u) &=\frac1{12} Nu   -\frac{11}{216} Nu^2 +  \frac1{12}{Nu^3}\left(\frac{103 N}{648}+\frac{2677}{1296}-\zeta_3\right)
+\mathcal O(u^4)\,,
\notag\\[2mm]
\gamma_\varphi(u) &=\frac{1}{6} u
-\frac{11(N+2)}{432} u^2 +\frac{u^3}6\left( -\frac{13N^2}{5184}+\frac{641 N}{2592} +\frac{2461}{1296}-\zeta_3\right)
+\mathcal O(u^4)\,,
\notag\\[2mm]
\beta(u) &= -2\epsilon u +\frac16(N+4) u^2 -\frac{11 N+119}{54} u^3 + \mathcal O(u^4)\equiv -2u(\epsilon+\bar\beta(u))
\end{align}
for the model~II. 
Note that for $N=2$ one finds $\gamma_\sigma=\gamma_\varphi$ as required by the symmetry. Both models have an
infrared (IR) stable fixed point below $d=6$ dimensions. The peculiar feature of these models is the absence of one-loop vertex correction. This greatly reduce\HL{s} the number of possible topologies for the higher-loop diagrams.  
\subsection{Leading twist operators: model $\text{I}$}
In model~I we consider the following operators: ``non-singlet'' operators (with respect to the symmetry group  $\mathbb{Z}_3$, $\varphi\to e^{\pm i2\pi/3}\varphi$)
\begin{align}
\mathcal O^{{\text{ns}}}_{s}=\varphi \partial_{\mu_1}\ldots\partial_{\mu_{s-1}}\varphi -\text{traces},
\end{align}
and ``singlet'' operators
\begin{align}
\mathcal O^{\pm}_{s}=\bar \varphi \partial_{\mu_1}\ldots\partial_{\mu_{s-1}}\varphi \mp (\varphi\leftrightarrow \bar\varphi) -\text{traces}.
\end{align}
Note here that the spin of the operator is equal to $s-1$. The  non-singlet operators are defined for odd $s$ and the singlet operators --
$\mathcal O^{+}_s$ and $\mathcal O^{-}_s$ -- for even and odd $s$, respectively.  We {use} such a definition because for this choice of the
parameter $s$  the mass correction formula, Eq.~\eqref{deltam2}, keeps its form in $d=6$ dimensions.

For the non-singlet operators the anomalous dimensions take the form
\begin{align}
\gamma^{{\text{ns}}}(s) &=2\gamma_\varphi -\frac{2 u^2}{s^2(s+1)^2} + \mathcal O(u^3),
\end{align}
and  the corresponding mass correction $\delta m^2(s)=2\Big(s+\bar\beta+\frac12\gamma^{{\text{ns}}}(s)\Big)\gamma^{{\text{ns}}}(s)$
diverges at $s\to0$.

For the singlet operators $\mathcal O^{\pm}_{s}$ we obtain
\begin{align}
\gamma^{\pm}(s) &= 2\gamma_\varphi \pm \frac{2u}{s(s+1)} \pm \frac{u^2}{3s(s+1)}\left(S_1-\frac1{2s(s+1)} - 4 +\frac{2s+1}{s(s+1)}\right)
\notag\\
&\quad
+\frac{2u^2}{s^2(s+1)^2}\left(1-\frac{2s+1}{s(s+1)}\right)
\notag\\
&\quad
+ \dfrac{u^3}{3s^2 (s + 1)^2}\Biggl\{12S_3 + 24S_{1, -2} - 12S_{-3} - S_{2} - 12\left(2 + \dfrac{1}{s(s + 1)}\right)S_{-2} \notag \\
&\quad +2\left(7 - \dfrac{1 + 2s}{s(s + 1)}\right)S_1 - 36\zeta_3 - \dfrac{24\left(S_{-2} + 1\right)}{(s- 1)(s + 2)}
+ \dfrac{6s - 3}{2s^2(s + 1)^2} + \dfrac{20s - 1}{s(s + 1)}- 7\Biggr\}
\notag \\
&\quad \pm  \dfrac{u^3}{36s(s + 1)}\Biggl\{2S_2 + S_1^2 - 144\left(
\dfrac{1}{s^2(s+ 1)^2} + \dfrac{1}{s(s + 1)}\right)S_{-2}
\notag \\
&\quad
- 72\zeta_3 + \left(\dfrac{1 + 4s}{s(s + 1)} - \dfrac{143}{3}\right)S_1 + \dfrac{144\big(S_{-2} + 1\big)}{(s - 1)(s + 2)}
+ \dfrac{144}{s^4(s + 1)^4} + 144\dfrac{2 - 3s}{s^3(s + 1)^3}
\notag \\
&\quad
+\dfrac{289 - 8s}{2s^2(s + 1)^2} - \dfrac{538 + 286s}{3s(s + 1)} + \dfrac{4927}{24}\Biggr\} + O(u^4).
\end{align}
Note that the singularity at $s = 1$ is spurious. We have checked that the anomalous dimensions have the reciprocity respecting form and that the anomalous dimension of the stress-energy tensor
vanishes, $\gamma_-(3)=0$.
One can easily find that the correction $\delta m^2_+(s)$ is regular at $s=0$,
\begin{align}
\delta m^2_+(0) = 4u + \dfrac{41}{18}u^2 + \left(-\frac{16}{3}\zeta_3 -\frac{44}{5}\zeta_2^2 + 8\zeta_2 + \frac{19721}{1296}\right)u^3 + O(u^4),
\end{align}
while $\delta m^2_-(s) \simeq 8u^3(2 - 3\zeta_3)/s $. Thus we see  that  the $\delta m^2(s)$ correction is finite at $s=0$ only for the
anomalous dimensions continued from {\it even} $s$ (which correspond to odd spin operators in the cubic model).

\subsection{Leading twist operators: model $\text{II}$\label{phi3-model-ii}}
In this Subsection we consider {only operators of odd spin (which corresponds to the even $s$)}. Let us start with
operators constructed of  the holomorphic fields, $\sigma$ and $\varphi$,
\begin{align}
\mathcal{O}_s^{a} & =\sigma\partial_{\mu_1}\ldots\partial_{\mu_{s-1}} \varphi^a -\text{traces},
\notag\\
\mathcal{O}_s^{[ab]} &=\varphi^{a}\partial_{\mu_1}\ldots\partial_{\mu_{s-1}} \varphi^b -(a\leftrightarrow b)-\text{traces},
\end{align}
The loop corrections for these operators  vanish in the first two orders of perturbation theory and the anomalous dimensions take the form\footnote{Note that for
$N=2$ these anomalous dimensions coincide.}
\begin{align}
\gamma^a(s)& =\gamma_\sigma+\gamma_\varphi -\frac{u^3}6(25N-2) \frac{S_1(s)}{s^2(s+1)^2} + O(u^4)\,,
\notag\\
\gamma^{[ab]}(s)& =2\gamma_\varphi +\frac{u^3}3(N-26) \frac{S_1(s)}{s^2(s+1)^2} + O(u^4)\,.
\end{align}
Since $S_1(s) = \zeta_2s + O(s^2)$, the corresponding mass corrections are finite at $s\to 0$.
\vskip 3mm

Traceless and symmetric operators constructed from holomorphic and anti-holomorphic  fields include two scalar operators,
\begin{align}
\label{phi3-singlet-2}
  \overline{\mathcal{O}}_s =\bar\varphi\partial_{\mu_1}\ldots\partial_{\mu_{s-1}} \varphi, 
  &&
    \Sigma_s = \bar\sigma\partial_{\mu_1}\ldots\partial_{\mu_{s-1}} \sigma, 
\end{align}
which mix under renormalization, and several $O(N)$ irreducible  tensor operators\footnote{Note that $\overline{\mathcal{O}}_{2}^{[ab]}$
corresponds to the conserved $O(N)$ current.}
\begin{align}
    \overline{\mathcal{O}}_s^{a} & =\bar\sigma\partial_{\mu_1}\ldots\partial_{\mu_{s-1}} \varphi^a,
\notag\\
   \overline{\mathcal{O}}_s^{{ab}} &=\bar\varphi^{a}\partial_{\mu_1}\ldots\partial_{\mu_{s-1}} \varphi^b  - N^{-1}\delta^{ab}\overline{\mathcal{O}}_s + (a\leftrightarrow b)
,
   \notag\\
   \overline{\mathcal{O}}_s^{[ab]} &=\bar\varphi^{a}\partial_{\mu_1}\ldots\partial_{\mu_{s-1}} \varphi^b -(a\leftrightarrow b). 
\label{phi3-nonsinglet-1}
\end{align}
The expressions for  the anomalous dimensions are quite long  and can be found in the Appendix~\ref{app:thre-loop-res}. Here we present only
the corresponding small $s$ expansion
\begin{align}
\bar{\gamma}^a(s) &= u\left\{\dfrac{2}{s} + \dfrac{N - 22}{12} + 2s - 2s^2 + 2s^3 + O(s^4)\right\}
+ u^2\Biggl\{-\dfrac{2}{s^3} + \dfrac{13}{3s^2} - \dfrac{98 + 5N}{18s}
\notag \\
&\quad
+ \dfrac{(N + 2)}{6}\zeta_2 + \dfrac{15N + 1178}{432} + \left(-\dfrac{N + 2}{6}\zeta_3
- \dfrac{N + 2}{6}\zeta_2 + \dfrac{N + 34}{18}\right)s
+ O(s^2)\Biggr\} \notag \\ &\quad
+ u^3\Biggl\{\dfrac{4}{s^5} - \dfrac{13}{s^4} + \dfrac{10N + 437}{18s^3}
- \left(\dfrac{N + 2}{3}\zeta_2 + \dfrac{189N + 6938}{216}\right)\dfrac{1}{s^2} - \Biggl(4\zeta_3
\notag \\
&\quad
+ \dfrac{22(N + 1)}{5}\zeta_2^2 - \dfrac{169N + 194}{36}\zeta_2 + \dfrac{3N^2 - 712N - 25309}{648}\Biggr)\dfrac{1}{s} + O(s^0)\Biggr\} + O(u^4),
\end{align}
and
\begin{align}
\bar{\gamma}^{(ab)}(s) &= u\left\{\dfrac{2}{s} - \dfrac{5}{3} + 2s - 2s^2 + 2s^3 + O(s^4)\right\}
+ u^2\Biggl\{-\dfrac{2}{s^3} + \dfrac{N + 24}{6s^2} - \dfrac{2(2N + 23)}{9s}  \notag
\\
&\quad
+ \dfrac{2}{3}\zeta_2 + \dfrac{49N + 506}{216} -\left(\dfrac{2}{3}\zeta_3 + \dfrac{2}{3}\zeta_2 + \dfrac{N - 20}{9}\right)s  + O(s^2)\Biggr\}
+ u^3\Biggl\{\dfrac{4}{s^5} - \dfrac{N + 24}{2s^4} \notag \\ &\quad  + \dfrac{N^2 + 112N + 1600}{72s^3}
 -\left(\dfrac{4}{3}\zeta_2 + \dfrac{4N^2 + 249N + 3144}{108}\right)\dfrac{1}{s^2} + \Biggl(-4\zeta_3 -
 \dfrac{66}{5}\zeta_2^2
 \notag \\
 &\quad
 + \dfrac{N + 264}{18}\zeta_2 + \dfrac{12N^2 + 1621N + 23431}{648}\Biggr)\dfrac{1}{s} + O(s^0)\Biggr\} + O(u^4),
\\
\bar{\gamma}^{[ab]}(s) &= u\left\{-\dfrac{2}{s} + \dfrac{7}{3} - 2s + 2s^2 - 2s^3 + O(s^4)\right\}
+ u^2\Biggl\{-\dfrac{2}{s^3} - \dfrac{N - 24}{6s^2} + \dfrac{2(2N  -13)}{9s}
\notag \\
&\quad
-\dfrac{2}{3}\zeta_2- \dfrac{71N - 314}{216} + \left(\dfrac{2}{3}\zeta_3 + \dfrac{2}{3}\zeta_2 + \dfrac{N + 16}{9}\right)s
+ O(s^2)\Biggr\} + u^3\Biggl\{-\dfrac{4}{s^5} - \dfrac{N - 24}{2s^4}
\notag \\
&\quad
- \dfrac{N^2 -112N + 1280}{72s^3}
+ \left(-\dfrac{4}{3}\zeta_2 + \dfrac{4N^2 - 207N + 2472}{108}\right)\dfrac{1}{s^2} + \Biggl(4\zeta_3 - \dfrac{22}{5}\zeta_2^2
\notag \\
&\quad
-\dfrac{N - 120}{18}\zeta_2 - \dfrac{12N^2 - 683N + 21127}{648}\Biggr)\dfrac{1}{s} + O(s^0)\Biggr\} + O(u^4).
\end{align}
Again, the  mass corrections for the operators under consideration are finite in the limit $s\to 0$. Namely,
\begin{align}
\delta m^{2}_a(0) &= 4u {-} \dfrac{N^2 {+} 88N - 244}{144}u^2 + \Biggl(-\dfrac{2(N + 14)}{3}\zeta_3
- \dfrac{44(N + 1)}{5}\zeta_2^2 \notag \\ &\quad  + 8(N + 1)\zeta_2 + \dfrac{10N^2 + 1255N + 29422}{1296}\Biggr)u^3 + O(u^4),
\notag\\
\delta m^{2}_{(ab)}(0) &= 4u + \dfrac{30 - 11N}{18}u^2 + \Biggl(-\dfrac{32}{3}\zeta_3 - \dfrac{132}{5}\zeta_2^2 + 24\zeta_2
\notag \\ &\quad - \dfrac{N^2 - 618N - 30740}{1296}\Biggr)u^3 + O(u^4),
\notag \\
\delta m^{2}_{[ab]}(0) &= -4u + \dfrac{9N + 110}{18}u^2 + \Biggl(\dfrac{16}{3}\zeta_3
- \dfrac{44}{5}\zeta_2^2 + 8\zeta_2 \notag \\ &\quad  + \dfrac{23N^2 - 2414N - 33444}{1296}\Biggr)u^3 + O(u^4).
\end{align}

For the singlet operators we present the expansion for the  anomalous dimension  matrix
\begin{align}
\label{matrix-expansion}
\setlength{\arraycolsep}{0.2cm}
\widehat{\gamma}(s) &=
2u\Biggl\{\begin{pmatrix}
    1 && N \\
    1 && 0
\end{pmatrix}\dfrac{1}{s} - \begin{pmatrix}
    \frac{5}{6} && N \\
    1 && -\frac{N}{12}
\end{pmatrix} + \begin{pmatrix}
    1 && N \\
    1 && 0
\end{pmatrix}s - \begin{pmatrix}
    1 && N \\
    1 && 0
\end{pmatrix}s^2 + \begin{pmatrix}
    1 && N \\
    1 && 0
\end{pmatrix}s^3 + O(s^4)\Biggr\} \notag \\ &\quad
+ u^2\Biggl\{-\begin{pmatrix}
    2N + 2 && 2N \\
    2 &&  2N
\end{pmatrix}\dfrac{1}{s^3} +
\setlength{\arraycolsep}{0.2cm}
\left(
\begin{array}{cc}
 \frac{25}{6}N + 4 & \frac{13}{3}N \\
 \frac{13}{3} & 4N \\
\end{array}
\right)\dfrac{1}{s^2} -
\left(
\begin{array}{cc}
 \frac{40}{9}N + \frac{46}{9} & 6N\\
 \frac{5}{9}N + \frac{44}{9} & 4N \\
\end{array}
\right)\dfrac{1}{s} \notag \\ &\quad
+\setlength{\arraycolsep}{0.2cm}\left(
\begin{array}{cc}
    \frac{2}{3} & \frac{2}{3}N \\
    \frac{1}{3}N & 0
\end{array}
\right)\zeta_2
+ \left(
\begin{array}{cc}
\frac{481}{216}N + \frac{253}{108} &  3N \\
\frac{2}{9}N + \frac{23}{9} & \frac{205}{108}N \\
\end{array}
\right) + \Biggl[ -\left(
\begin{array}{cc}
  \frac{2}{3}  & \frac{2}{3}N\\
 \frac{N}{3}  & 0 \\
\end{array}
\right)\zeta_3\notag \\ &\quad - \setlength{\arraycolsep}{0.2cm}\left(
\begin{array}{cc}
  \frac{2}{3} & \frac{2}{3}N\\
 \frac{1}{3}N & 0 \\
\end{array}
\right)\zeta_2 + \left(
\begin{array}{cc}
 \frac{17}{9}N + \frac{20}{9} & 2N\\
 \frac{1}{9}N + \frac{16}{9} & 2N \\
\end{array}
\right)\Biggr]s
+ O(s^2)\Biggr\}
\notag \\ &\quad + \setlength{\arraycolsep}{0.2cm} u^3\left(
\begin{array}{cc}
 \gamma_{\varphi\varphi}^{(3)}(s) & \gamma_{\varphi\sigma}^{(3)}(s) \\
 \gamma_{\sigma\varphi}^{(3)}(s) & \gamma_{\sigma\sigma}^{(3)}(s) \\
\end{array}
\right) + O(u^4),
\end{align}
where
\begin{align}
    \gamma^{(3)}_{\varphi\varphi}(s) &= 4\frac{2N + 1}{s^5} - \frac{N^2 + 151N + 72}{6s^4} + \frac{65 N^2+3248 N+1600}{72 s^3}
     -\Biggl(\left(N^2 + 2N + 4\right)\zeta_2
     \notag \\
     &\quad  + \frac{198N^2 +5837N + 3144}{36}\Biggr)\dfrac{1}{3s^2} + \Biggl(\frac{(N-6)(N+4)}{6}\zeta_3 - \frac{66}{5}\zeta_2^2
     \notag \\ &\quad  + \frac{17 N^2+15 N+264}{18} \zeta_2 + \frac{1116 N^2+33973 N+23431}{648}\Biggr)\dfrac{1}{s}  + O(s^0),
    \\
    \gamma^{(3)}_{\varphi\sigma}(s) &= N\Biggl\{\frac{4 (N+1)}{s^5} - \frac{73N +76}{6 s^4} +\frac{1475 N+1758}{72 s^3}
     - \Biggl( 4\zeta_2 + \frac{5297N + 7090}{72}\Biggr)\dfrac{1}{3s^2}
     \notag \\
     &\quad \Biggl(-4\zeta_3 - \frac{22}{5}\zeta_2^2 + \frac{61}{9}\zeta_2  + 5\frac{4283 N+6854}{864} \Biggr)\dfrac{1}{s} + O(s^0)\Biggr\},
    \\
    \gamma^{(3)}_{\sigma\varphi}(s) &= \frac{4 (N+1)}{s^5} - \frac{73N + 76}{6s^4} +\frac{1533 N+1642}{72 s^3}
     - \Biggl((N+ 2)\zeta_2 + \frac{5517N +6650}{72}\Biggr)\dfrac{1}{3s^2}
      \notag \\
      &\quad
    \Biggl(- \frac{N + 22}{6}\zeta_3 - \frac{22}{5}\zeta_2^2 + \frac{4N + 53}{9}\zeta_2
     -\frac{36 N^2 - 66107N - 99230}{2592}\Biggr)\dfrac{1}{s} + O(s^0),
    \\
    \gamma^{(3)}_{\sigma\sigma}(s) &=
    N\Biggl\{\frac{4}{s^5} +\frac{N-80}{6 s^4}  +\frac{2 (N+112)}{9 s^3} - \Biggl((N + 2)\zeta_2 - \frac{13N - 1718}{18}\Biggr)\dfrac{1}{3s^2}
     \notag \\ &\quad \Biggl(- \frac{N - 2}{6}\zeta_3
     - \frac{22}{5}\zeta_2^2 + \frac{7N+106}{18}\zeta_2 - 22\frac{N-38}{27}\Biggr)\dfrac{1}{s} + O(s^0)\Biggr\}.
\end{align}
The mass correction matrix reads
\begin{equation}
    \delta \widehat{m}^2(s) = 2\left(\Big(s + \bar{\beta}(u)\Big)\mathds{1} + \dfrac{1}{2}\widehat{\gamma}(s)\right)\widehat{\gamma}(s),
\end{equation}
where $\mathds{1}$ is an identity matrix. It is regular at $s=0$ in the first two orders of perturbation theory,
\begin{align}
\label{delta-matrix}
    \delta \widehat{m}^2(0) &= 4u\begin{pmatrix}
        1 && N \\
        1 && 0
    \end{pmatrix}  + \dfrac{u^2}{18} 
    \begin{pmatrix}
        61N + 30 && 12N \\
        -20N + 52 && 70N
    \end{pmatrix}  +O(u^3)\,.
\end{align}
The three-loop matrix elements retain $1/s^2$ and  $1/s$ poles which however cancel in the expressions for the sum and product of the
mass eigenvalues, $\delta m_k^2$, $k=1,2$.
Namely
\begin{align}
\delta m^2_1+\delta m_2^2  &=  \operatorname{tr}\Big[\delta \widehat{m}^2\Big](0)
= 4u + \dfrac{131 N+30}{18}u^2 + \Biggl(- \frac{4\left(N^2+2 N+8\right)}{3}\zeta_3
    -\frac{44}{5}(N + 3)\zeta_2^2
    \notag \\
    &\quad
  + 8(N + 3)\zeta_2   - \frac{1131 N^2 - 25950N - 30740}{1296}\Biggr)u^3 + O(u^4),
\\
 \delta m^2_1\cdot\delta m_2^2 &=  \operatorname{det}\Big[\delta \widehat{m}^2\Big](0) = N\Biggl\{-16u^2
 + \dfrac{4(10N + 3)}{9}u^3 + \Biggl(\frac{224}{3}\zeta_3 +\frac{176}{5}\zeta_2^2 -32\zeta_2
 \notag \\
 &\quad -\frac{28 N^2+4421 N+12504}{81}\Biggr)u^4 + O(u^5)\Biggr\}.
\end{align}
This implies regularity of $\delta m^2_k$ at $s=0$.  Note also that
\begin{align}
\delta m_{1,2}^2(0)=2u\left(1\pm \sqrt{1 + 4N}\right)+O(u^2),
\end{align}
so $\delta m_1^2(0) > 0$ and $\delta m^2_2(0) < 0$ for $N > 0$. 


\section{Gross-Neveu-Yukawa model\label{sect:GNY}}
In this Section, we consider mass corrections in the GNY model~\cite{ZinnJustin:1989mi},
\begin{equation}
\label{GNY-action-def}
    S = \int d^dx \left\{\bar{q}\slashed{\partial}q + \dfrac{1}{2}(\partial\sigma)^2 + g\bar{q}\sigma q
     + \dfrac{\bar{\lambda}}{4!}\sigma^4\right\},
\end{equation}
where $q(x)$ is an $N$ component quark  field 
and $\sigma(x)$ is a scalar field.
The RG functions in this model are known with four loop accuracy~\cite{Zerf:2017zqi}
\begin{align}
\label{GNY-beta-expr}
    \beta_u(u, \lambda) &= -2u\epsilon +  2(3 + 2N)u^2 + O(u^3, \lambda^3), \nonumber\\
    \beta_\lambda(u,\lambda) &= -2\lambda\epsilon + 3\lambda^2 + 8Nu\lambda - 48Nu^2 + O(u^3, \lambda^3),
\\
    \gamma_{q}(u, \lambda) &= \dfrac{u}{2} - \dfrac{u^2}{8}\left(12N + 1\right) + O(u^3, \lambda^3), \notag \\
    \gamma_{\sigma}(u, \lambda)
    &= 2Nu + \dfrac{1}{12}\lambda^2 - 5Nu^2 + O(u^3, \lambda^3),
\end{align}
where
\begin{align}
    u = \dfrac{g^2}{(4\pi)^2}, && \lambda = \dfrac{\bar{\lambda}}{(4\pi)^2}.
\end{align}
%
The model possesses an IR-stable fixed point (for details see Refs.~\cite{ZinnJustin:1989mi,Vasilev:2004yr,Zerf:2017zqi}) 
\begin{align}
\label{GNY-crit-point}
    u_* &= u_1\epsilon + u_2\epsilon^2 + \ldots = \dfrac{\epsilon}{3 + 2N} \notag \\ &\quad - \dfrac{8N^2 - 1032N - 441 - \left(4N + 66\right)\sqrt{4N^2 + 132N +9}}{108(3 + 2N)^3}  \epsilon^2 +  O(\epsilon^3), \notag \\[2mm]
    \lambda_* &= \lambda_1\epsilon + \lambda_2\epsilon^2 + \ldots = \dfrac{3 - 2N + \sqrt{4N^2 + 132N + 9}}{9 + 6N}\epsilon + O(\epsilon^2). 
\end{align}

We consider the following twist-two operators, namely the flavor non-singlet operators
\begin{equation}
    \mathcal{O}_s^{q, a} = \bar{q}_i\gamma_{\mu_1}\partial_{\mu_2}\ldots\partial_{\mu_s} t^{a}_{ij}q_j,
\end{equation}
where $t^{a}_{ik}$ ($a = 1, \ldots, N^2- 1$) are the generators of $SU(N)$ flavor group, 
and the flavor singlet operators\footnote{We choose such a normalization for the quark operator for later convenience.}
\begin{align}\label{GNYsinglet}
    \mathcal{O}_s^{q} = 
        -s\,\bar{q}_i\gamma_{\mu_1}\partial_{\mu_2}\ldots\partial_{\mu_s}q_i, &&
    \mathcal{O}_s^{\sigma} = 
           \sigma\partial_{\mu_1}\ldots\partial_{\mu_s}\sigma.
\end{align}
Note that the operator $\mathcal{O}_{s}^{\sigma}$ does not exist for odd $s$.

We calculate anomalous dimensions of twist-two operators in the two-loop approximation. For the non-singlet operator we obtain
\begin{align}
\label{GNY-nonsinglet}
    \gamma^q(s) &= 2\gamma_q - \dfrac{2u}{s(s + 1)}
    - \dfrac{2u^2}{s(s + 1)}\Biggl(3S_1 + \dfrac{1 + 2s}{s^2(s + 1)^2} + \dfrac{1 + 5s + 2N(1 + 2s)}{s(1 + s)}
    \notag \\
    &\quad - 8 - 6N + (1 - (-1)^s)\left(1 - \dfrac{1}{s(s + 1)}\right)\Biggr) + O(u^3, \lambda^3).
\end{align}
One can readily check that $\gamma^{q}(s = 1) = 0$, as it corresponds to the conserved current.
The singlet anomalous dimension matrix, defined for {\it even} $s$ only, takes the form
\begin{align}
\label{GNY-matrix}
    \widehat{\gamma}(s) = \begin{pmatrix}
        2\gamma_{q} && 0 \\
        0 && 2\gamma_{\sigma}
    \end{pmatrix} - 2u\begin{pmatrix}
        \frac{1}{s(s + 1)} && 4N \\
        \frac{1}
        {s(s + 1)} && 0
    \end{pmatrix} + \begin{pmatrix}
        \gamma^{(2)}_{qq}(s) && \gamma^{(2)}_{q\sigma}(s) \\
        \gamma^{(2)}_{\sigma q}(s) && \gamma^{(2)}_{\sigma\sigma}(s)
    \end{pmatrix} + O(u^3, \lambda^3),
\end{align}
where
\begin{align}
    \gamma_{qq}^{(2)}(s) &=  - \dfrac{2u^2}{s(s + 1)}\Biggl(3S_1 + \dfrac{1 + 2s}{s^2(s + 1)^2} + \dfrac{1 + 5s + 2N(3 + 2s)}{s(1 + s)}
    - 8 - 10N\Biggr),
    \notag \\
    \gamma_{q\sigma}^{(2)}(s) &= -  8N u^2 \Bigg(3S_1 + \dfrac{1 + 2s}{s^2(s + 1)^2} + \dfrac{3}{2s(s + 1)} - 7\Bigg),
    \notag \\
    \gamma_{\sigma q}^{(2)}(s) &= -
    \frac{4u^2}{s(s + 1)}\Biggl((1 + 2N)S_1 + \dfrac{1 + 2s}{2s^2(s + 1)^2} + \dfrac{7 + 2(5 + 4N)s}{4s(s+1)} - 4N - \frac{5}{2}\Biggr),
    \notag \\
    \gamma_{\sigma\sigma}^{(2)}(s) &= 
   -\frac{8N u^2} {s(s+ 1)}\Biggl(\dfrac{1 + 2s}{s(s + 1)} - 2\Biggr) - \dfrac{\lambda^2}{s(s + 1)}.
\end{align}
Projection of the matrix $\widehat{\gamma}$ with the vector $v^T=(1,2)$ shows that
\begin{equation}
  v^T \cdot \widehat{\gamma}(s = 2) = 
 \begin{pmatrix}
        1, &&  2
      \end{pmatrix}
  \cdot \widehat{\gamma}(s = 2) = 0,
\end{equation}
which corresponds to the conserved stress-energy tensor.

We consider the mass correction~\eqref{deltam2} at the critical point given in Eq.~\eqref{GNY-crit-point}. 
For the non-singlet operators we define
\begin{equation}
    \delta m_{\pm}^2(s) = 2\left(s - \epsilon + \dfrac{1}{2}\gamma^{q}_{*}(s)\right)\gamma^{q}_{*}(s),
\end{equation}
where $\pm$  functions are defined for the even (odd) values of $s$  and $\gamma_{*}^{q}(s)\equiv \gamma^q(s,u_\ast,\lambda_\ast)$.
The function $\delta m_{+}^{2}(s)$ is regular at $s = 0$
\begin{equation}
    \delta m^{2}_{+}(0) = - 4u_1\epsilon + \left((24N + 37)u_1^2 - 6u_1 - 4u_2\right)\epsilon^2 + O(\epsilon^3),
\end{equation}
while $\delta m^2_-$ is singular, $\delta m^2_{-}(s) \simeq 8u_1^2/s$.

For the singlet operators we define the mass correction as
\begin{equation}
\delta\widehat{m}^2(s) = 2\left((s - \epsilon)\mathds{1} + \dfrac{1}{2}\widehat{\gamma}_*(s)\right)\widehat{\gamma}_*(s).
\end{equation}
The eigenvalues of the matrix $\delta\widehat{m}^2(s)$ are finite at $s\to 0$,
\begin{align}\label{meigenvalues}
    \delta\widehat{m}_1^2(0) &= {-}4u_1\epsilon + \Big((16N - 6) u_1 -(32N^2 + 24N - 37) u_1^2  - 4 u_2\Big)\epsilon^2 + O(\epsilon^3),
    \notag\\[2mm]
    \delta\widehat{m}_2^2(0) &= \Big( 48 N(N+2)  u_1^2 - 24 N u_1 - 2\lambda_1^2 \Big)\epsilon^2 + O(\epsilon^3).
\end{align}
Moreover, with the help of the relations for $u_*$ and $\lambda_*$ in Eq.~\eqref{GNY-crit-point}, one can show that 
\begin{equation}\label{delta2minus}
   \delta\widehat{m}_2^2(0) = \left(-2\epsilon + \gamma^*_{\sigma^2}\right)\gamma^{*}_{\sigma^2}
   = \left(\lambda_1 + 4 N u_1\right)\left(\lambda_1 + 4 N u_1 - 2\right)\epsilon^2 + O(\epsilon^3),
\end{equation}
where $\gamma_{\sigma^2}^{*}=(\lambda_1+4Nu_1)\,\epsilon +O(\epsilon^2)$ is the anomalous dimension of the operator $\sigma^2$. 

Using analyticity of Eq.~\eqref{meigenvalues} we define two trajectories around $s = 0$ (note that $\epsilon \ll 1$)
\begin{equation}
    \gamma_{1,2}(s) = \epsilon - s + \sqrt{(\epsilon - s)^2 + \delta \widehat{m}^2_{1,2}(s)}.
\end{equation}
For large $s$ they correspond to the eigenvalues of the anomalous dimension matrix~\eqref{GNY-matrix} and remain regular around $s = 0$. 
Branching points of these trajectories are defined as 
\begin{align}
    s^{\pm}_{1} = \pm 2\sqrt{\epsilon/(2N + 3)} + O(\epsilon), && s^{\pm}_{2} = \Biggl(\dfrac{9 - 30N}{9 + 6N} \pm a_N\Biggr)\epsilon + O(\epsilon^2),
\end{align}
where $a_N \in \mathbb{R}$ for $N \ge 1$. All of these points lie on the real axis.

\subsection{Gross-Neveu model at large $N$}\label{sect:largeN}
The GNY model in $d$ dimensions is critically equivalent to the Gross-Neveu model~\cite{Gross:1974jv}. 
The latter can be analyzed in the $1/N$ expansion framework.
The index $\eta$ determining the critical dimension of the basic fermion field\footnote{In this Subsection we use the
standard notation  $\mu\equiv d/2$.}
 ($\Delta_q=\mu-1+\eta/2$) is known with an accuracy
$1/N^3$~\cite{Vasiliev:1992wr,Gracey:1993kc} and the critical dimensions of the auxiliary $\sigma$ field ( $\Delta_\sigma= 1 + \gamma_\sigma$ )
 and the operator $\sigma^2$ (~$\Delta_{\sigma^2}=2+\gamma_{\sigma^2}$~) with
an accuracy  $1/N^2$ \cite{Gracey:1992cp,Vasiliev:1993pi,Gracey:1993kb}.

The singlet operators $\mathcal O^q_s$ and $\mathcal O^\sigma_s$ defined in Eq.~\eqref{GNYsinglet} have different canonical dimensions, $d-2+s$ and $2+s$,
and therefore do not mix under renormalization. 
The anomalous dimensions of the quark operator were calculated at order $1/N$ in~\cite{Muta:1976js} and order $1/N^2$ in~\cite{Manashov:2016uam}.
For the operator $\mathcal O^\sigma_s$ the anomalous dimensions are known only at order $1/N$ \cite{Giombi:2017rhm}. 
The corresponding $1/N$ results are quite compact,
\begin{subequations}
\label{qsigmaN}
\begin{align}
\gamma_\sigma(s) &=-\dfrac{\eta_1}{n}\frac{2(2\mu-1)}{(\mu-1)}\left(1
-\frac{\mu}{2\mu-1}\frac{\Gamma(\mu)}{\Gamma(s+\mu)}\frac{\Gamma(s+2-\mu)}{\Gamma(3-\mu)} \right) +O(1/n^2)\,,
\\
\gamma_q(s) &=\dfrac{\eta_1}{n}\left(1-
\frac{\mu(\mu-1)}{(s+\mu-1)(s+\mu-2)}\left(1+\frac{\Gamma(2\mu-1) \Gamma(s+1)}{(\mu-1)\Gamma(2\mu-3+s)}\right)\right)+O(1/n^2)\,,
\end{align}
\end{subequations}
where $n=N\times\tr\II\equiv 4N$ and the index $\eta_1$ is
\begin{align}
\eta_1&=-\frac{2\Gamma(2\mu-1)}{\Gamma(\mu+1)\Gamma(\mu)\Gamma(\mu-1)\Gamma(1-\mu)}\,.
\end{align}
Analyzing the singularities we see that the anomalous dimensions, $ \gamma_\sigma(s)$ and $\gamma_q(s)$, have  poles at $s=\mu-2$ and $s=2-\mu$, respectively. Next, the  leading order
anomalous dimensions are finite at $s=0$, however, as was noted in~\cite{Giombi:2017rhm},  $\gamma_\sigma(s=0)\neq \gamma_{\sigma^2}$.
Given the analogous  situation   with $\varphi^4$ model, one would expect  $1/s$ poles to appear at higher orders. This is indeed the case
for the quark anomalous dimension,
\begin{align}
\label{quark-singular}
\gamma_q(s)&= \Biggl(-\frac{\eta_1^2}{n^2}\frac{2\mu(2\mu - 1)(2\mu - 3)}{(\mu-2)^2}\times \frac1s +O(s^0)\Biggr) + O(1/n^3) \,.
\end{align}
Thus the anomalous dimensions become singular at the points, $s=\pm(2-\mu)$ and $s=0$, 
which, according to Ref.~\cite{Caron-Huot:2022eqs}, can be identified with the intersection points of the following trajectories defined by scaling dimensions:
\begin{itemize}
\item the quark trajectory $\Delta_q(s)$ intersects with its shadow $\widetilde \Delta_q(s) = d - \Delta_q(s)$ at $s=2-\mu$.
\item the trajectory $\Delta_\sigma(s)$ intersects with $\widetilde \Delta_\sigma(s)$ at $s=\mu-2$.
\item the trajectory $\Delta_q(s)$ intersects with $\widetilde \Delta_\sigma(s)$, and $\Delta_\sigma(s)$ intersects with $\widetilde \Delta_q(s)$ at $s=0$.
\end{itemize}
Note that in the $1/N$ framework all four intersection points are well separated since one has not to assume that $\epsilon$ is small. The natural choice of the dimension for the theory in $1/N$ expansion is $2 < d < 4$, so we consider these limits in what follows. 

In the first two cases, assuming the analyticity of the products $\Delta_q(s)\widetilde \Delta_q(s)$ and $\Delta_\sigma(s)\widetilde
\Delta_\sigma(s)$ at the corresponding intersection points $s_{q,\sigma}=\pm(2-\mu)$,
one derives that the
following combination of the anomalous dimensions
\begin{subequations}
\begin{align}
\label{GNm2q}
\delta m_q^2(s) &=2\left(s+\mu-2 +\frac12\gamma_q(s)\right)\gamma_q(s)\,,
\\
\delta m_\sigma^2(s) &=2\left(s+2-\mu +\frac12\gamma_\sigma(s)\right)\gamma_\sigma(s),
 \end{align}
\end{subequations}
are regular functions of $s$  at these points.
At order $1/N$ this follows immediately from Eq.~\eqref{qsigmaN} and 
for $\delta m_q^2(s)$ at order $1/N^2$ it can be  checked using the results of Ref.~\cite{Manashov:2016uam}.
\begin{align}
\label{GNY-n-delMs}
\delta m_q^2(s_{q}) &=-\frac{\eta_1}{n}\mu(\mu-1)\left(1+\frac{\Gamma(2\mu-1)\Gamma(3-\mu)}{\Gamma(\mu)}\right)  + O\left(1/n^2\right),
\notag\\
\delta m_\sigma^2(s_{\sigma}) &=\frac{\eta_1}{n} \frac{8\Gamma(\mu+1)}{\Gamma(2\mu-1)\Gamma(3-\mu)} +O\left(1/n^2\right).
\end{align}
Here we have tacitly assumed that the anomalous dimensions $\gamma_\sigma(s)$ and $\gamma_q(s)$ are determined by analytic continuations 
from {\it even} integer $s$.
However the mass correction $\delta m_q^2(s)$ is also regular at the point $s=2-\mu$ for the anomalous dimensions $\gamma_q(s)$ continued from {\it odd} spins 
as well as for the non-singlet anomalous dimensions of both signatures\footnote{Operators of positive and negative signature are understood as continuation from the even and odd values of $s$, respectively.}.
Thus in the vicinity of the points $s_{q,\sigma}$, the anomalous dimensions acquire square root branching points
\begin{align}
\label{eps-traj}
\gamma_{\alpha}(s)= s_\alpha - s + \sqrt{(s-s_\alpha)^2 + \delta m^2_{\alpha}(s)},
\end{align}
where $\alpha=q,\sigma$.
Using Eq.~\eqref{GNY-n-delMs} one finds that in the  region $2 < d < 4$ mass corrections have a different sign, $\delta m^2_{q}(s_q) < 0$, $\delta m^2_{\sigma}(s_\sigma) > 0$. 
Therefore branching points of the square root lie on the real axis for the $\gamma_q(s)$ and become imaginary for the $\gamma_{\sigma}(s)$. 
The trajectories of Eq.~\eqref{eps-traj} around the points $s = s_{\alpha}$ are shown in Figure~\ref{plot:eps-point}. 
Note the different type of behavior when the branching points lie on the real axis $(a)$ or have a non-zero imaginary part $(b)$. 
\begin{figure}[h]
\begin{minipage}[h]{0.47\linewidth}
\center{\includegraphics[width=1\linewidth]{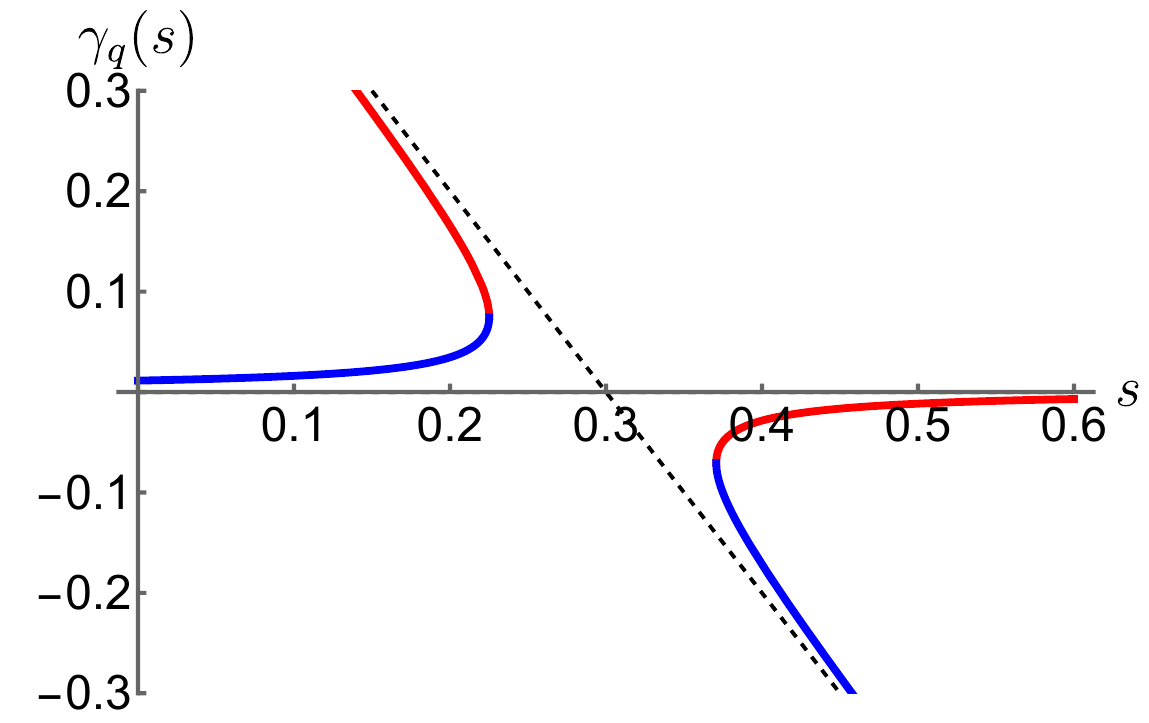}} (a) \\
\end{minipage}
\hfill
\begin{minipage}[h]{0.47\linewidth}
\center{\includegraphics[width=0.97\linewidth]{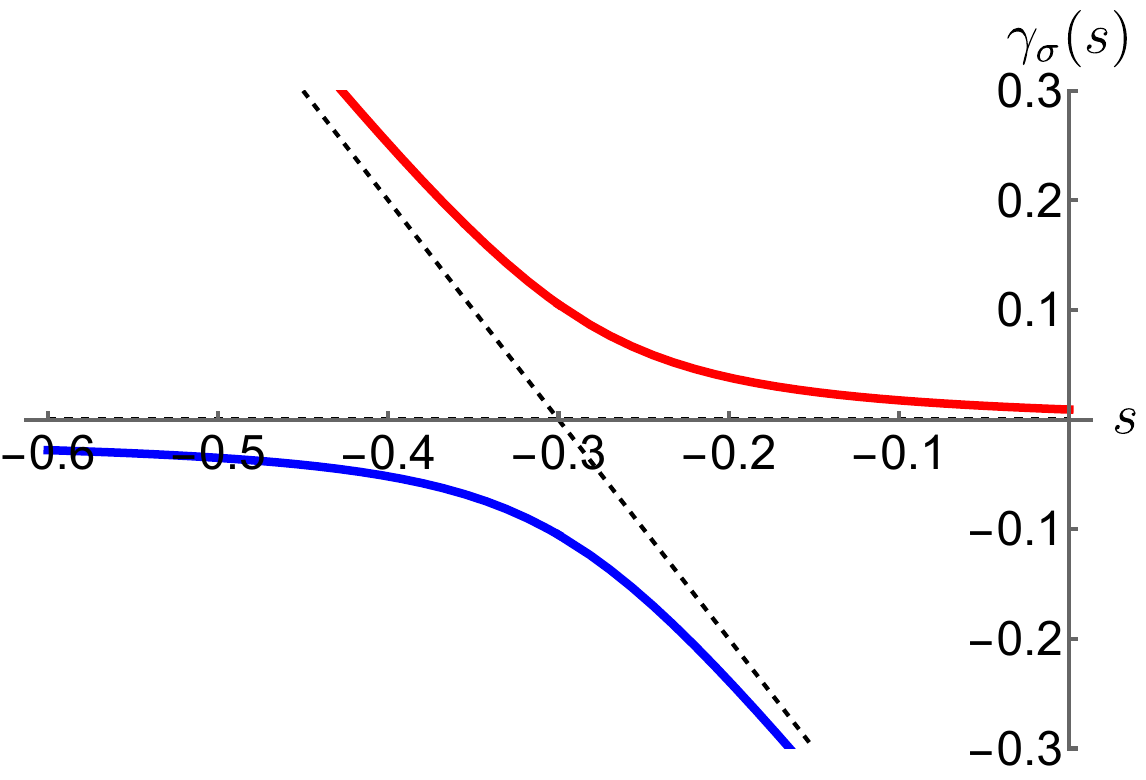}} \\(b)
\end{minipage}
\caption{\label{plot:eps-point} The anomalous dimensions defined as~\eqref{eps-traj} around the points $s = s_\alpha$. Graph $(a)$ shows $\gamma_{q}(s)$ and $(b)$ shows $\gamma_{\sigma}(s)$. Red and blue trajectories correspond to the two branches of the square root, while the dashed black line represents the trajectories in free theory $N \to \infty$. The plot is made for $\epsilon = 0.3$, $N = 40$.}
\end{figure}

At the point $s=0$ the quark trajectory intersects with the shadow $\sigma-$trajectory, and vice versa. The analyticity of the products
$\Delta_q(s)\widetilde \Delta_\sigma(s)$ and $\Delta_\sigma(s)\widetilde \Delta_q(s)$ is equivalent to the analyticity of the following
functions at $s=0$
\begin{subequations}
\label{s=0crossing}
\begin{align}
\omega_1(s)&=\frac12\left(\gamma_q(s)-\gamma_\sigma(s)\right)\,,
\\
\omega_2(s)&=s(\gamma_q(s)+\gamma_\sigma(s))+\gamma_q(s)\gamma_\sigma(s)\,.
\end{align}
\end{subequations}
The first of these equation states that poles of the anomalous dimensions of the operators $\mathcal O^q_s$ and   $\mathcal O^\sigma_s$ are
exactly the same in all orders of the $1/N$ expansion. In particular, we predict then that the 
singular structure of $\gamma_{\sigma}(s)$ coincides with Eq.~\eqref{quark-singular} in $1/N^2$ order. 
Unfortunately, the validity of general statement can be checked so far only in the order $1/N$, 
where both functions are
regular, since $\gamma_\sigma(s)$ is not known beyond this order.

Resolving Eq.~\eqref{s=0crossing} one derives for the anomalous dimensions in the vicinity of the point $s=0$
\begin{align}
\label{GNY-n-zero-trajs}
\gamma_{q,\sigma}(s)= \pm \omega_1(s) - s + \sqrt{s^2+\omega_2(s) +\omega_1^2(s)}.
\end{align}
Taking into account that
\begin{align}
\omega_1(0)&=-\frac{\eta_1}n \frac{ (2\mu-3)(\mu-1)}{\mu-2} + O\left(1/{n^2}\right),
\intertext{and}
\omega_2(0)&=-\left(\frac{\eta_1}n\right)^2 \frac{4(2\mu-1)}{\mu-2}  + O\left(1/{n^3}\right),
\end{align}
one finds that the anomalous dimension of the operator $\sigma^2$ is approached by the trajectory $\gamma_\sigma(s)$ on the first Riemann sheet,
\begin{align}
\lim_{s\to 0} \gamma_\sigma(s)=\gamma_{\sigma^2} =\frac{\eta_1}{n} 2(2\mu-1) + O\left(1/{n^2}\right)\,,
\end{align}
while
\begin{align}
\lim_{s\to 0} \gamma_q(s)=\frac{\eta_1}{n} \frac2{2-\mu} + O\left(1/{n^2}\right)\,.
\end{align}
The branching points of the trajectory~\eqref{GNY-n-zero-trajs} are located at
\begin{align}
\label{zero-bp}
    s^{\pm} = \dfrac{\eta_1}{n}\dfrac{(\mu - 1)(2\mu + 1) \pm 2\sqrt{\mu(2\mu - 1)(2\mu - 3)}}{\mu - 2} + O(1/n^2),
\end{align}
with the $1/N$ accuracy. In the region $2 < d < 3$ they lie on the real axis and acquire non-zero imaginary part in $3 < d < 4$. At the $d = 3$ the difference between two branching points loses, so one has to take into account the $1/N^2$ order. 
We plot the function $\gamma_{q}(s)$ around point $s = 0$ in Figure~\ref{plot:zero-point}. Note the change of the behavior for $d < 3$ $(a)$ and $d > 3$ $(b)$. The branching points~\eqref{zero-bp} are the same for quark and $\sigma$ anomalous dimensions, so the trajectory $\gamma_{\sigma}(s)$ looks the same as shown in Figure~\ref{plot:zero-point} modulo small corrections of the order $1/N$. 

\begin{figure}[h!]
\begin{minipage}[h!]{0.49\linewidth}
\center{\includegraphics[width=1\linewidth]{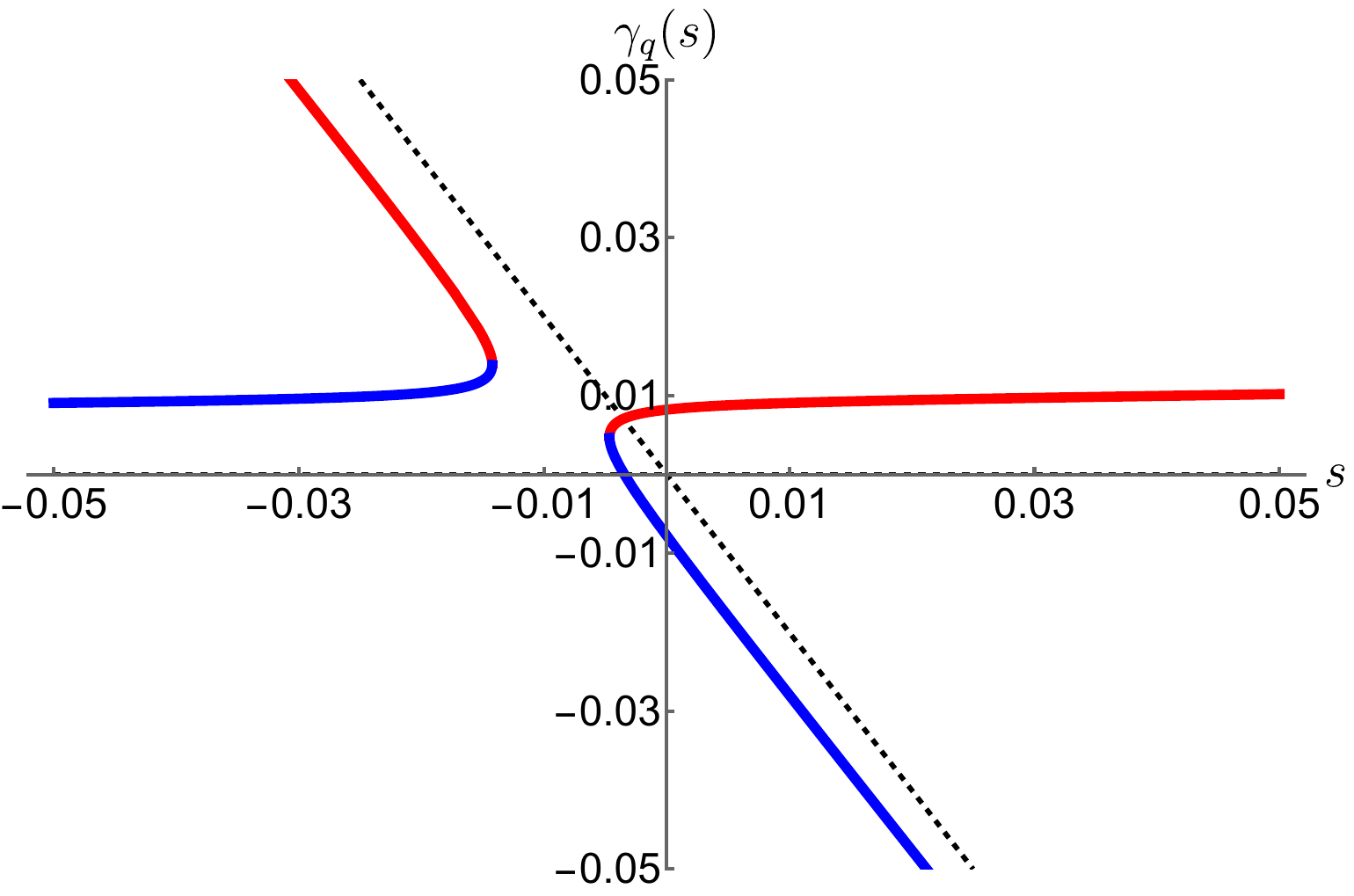}} (a) \\
\end{minipage}
\hfill
\begin{minipage}[h!]{0.47\linewidth}
\center{\includegraphics[width=1\linewidth]{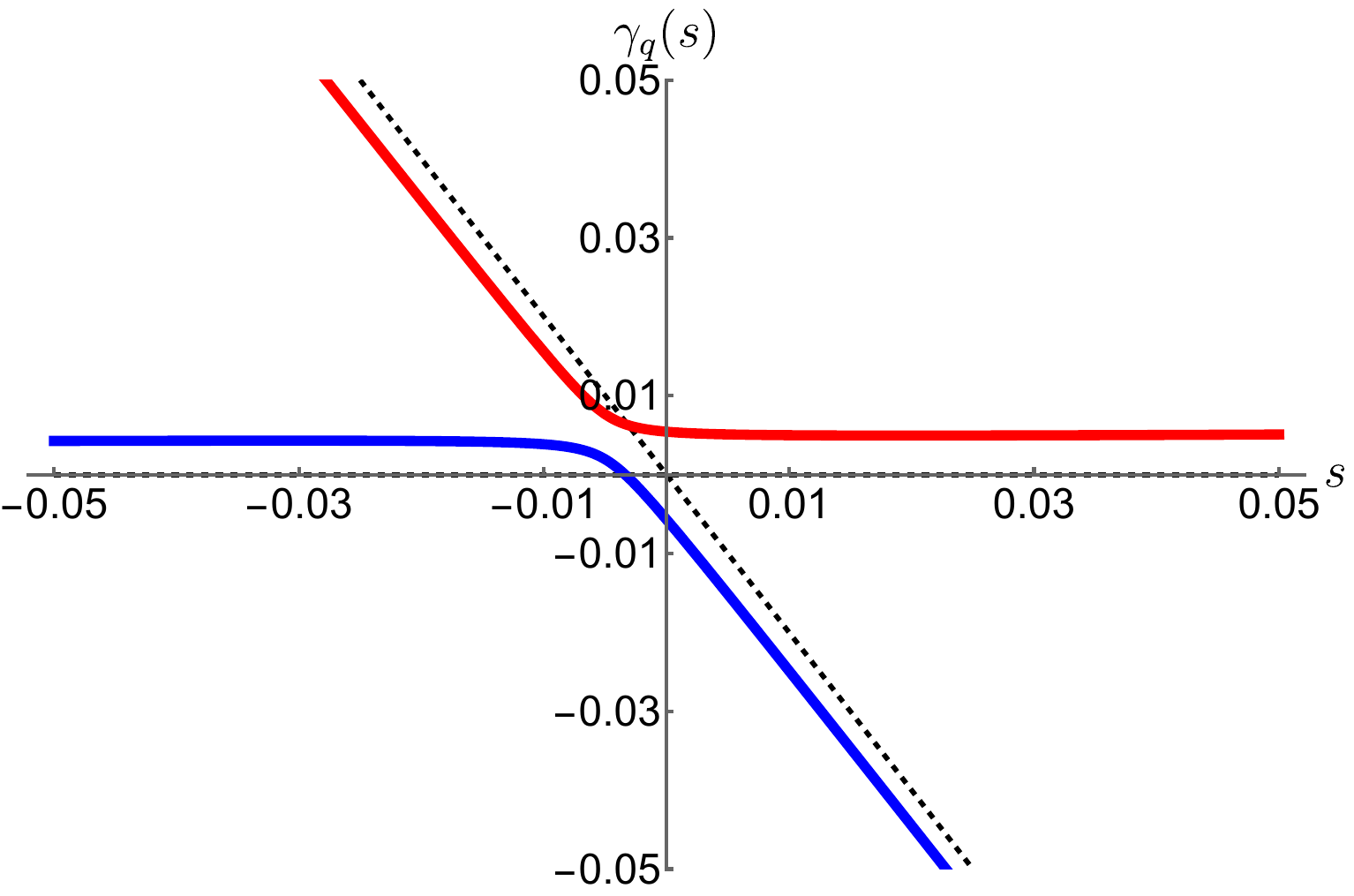}} \\(b)
\end{minipage}
\caption{\label{plot:zero-point} The anomalous dimension of the quark operator, $\gamma_{q}(s)$, defined as~\eqref{GNY-n-zero-trajs} around the point $s = 0$. The plot $(a)$ is made for $\epsilon = 0.45$, $N = 40$, while the plot $(b)$ is made for $\epsilon = 0.55$, $N = 40$. Red and blue trajectories correspond to the two branches of the square root, while the dashed black line represents the trajectories in free theory $N \to \infty$.}
\end{figure}

Taking into account information about anomalous dimensions around points $s = \pm(2- \mu)$ and $s = 0$, mentioned above, we can now plot the approximate graph of the 
scaling dimensions of operators $\mathcal{O}^q_s$ and $\mathcal{O}^{\sigma}_s$. In Figure~\ref{plot:full} the four scaling dimensions, 
$\Delta_q(s)$, $\Delta_s(s)$ and their shadows, $\widetilde{\Delta}_s(s)$ and $\widetilde{\Delta}_q(s)$, are shown in the axes $\Delta(s) - d/2$ and $s$. Unlike the results from the naive perturbation theory, resummation, done by the using~\eqref{eps-traj} and~\eqref{GNY-n-zero-trajs}, gives us the analytical behavior around points $s =\pm(2 - \mu)$ and $s = 0$. 
\begin{figure}[h!]
\center{\includegraphics[width=0.8\linewidth]{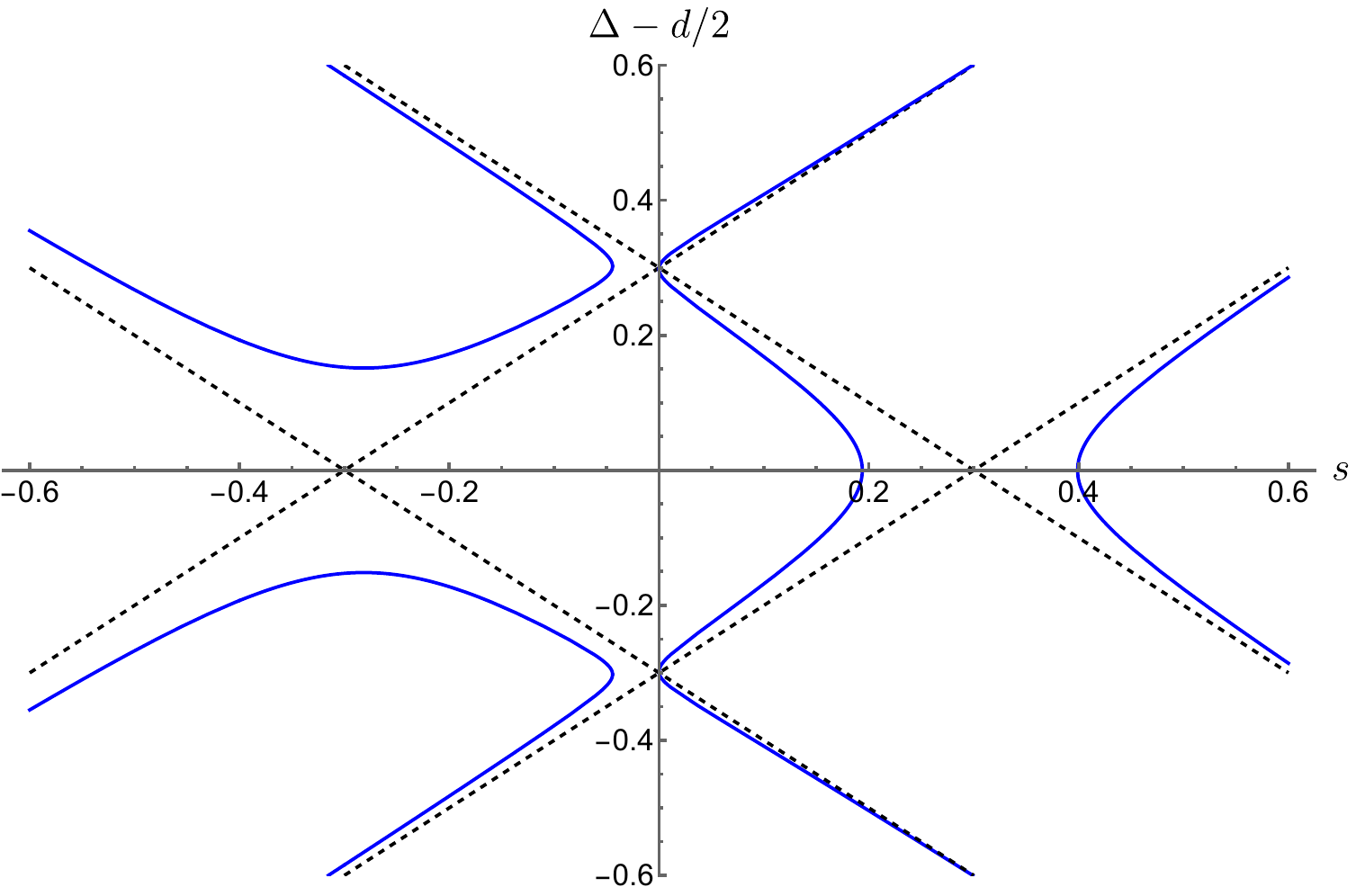}}
\caption{\label{plot:full}Blue solid line represents the scaling dimensions of quark and $\sigma$ operators and their shadows. Black dashed line shows the free theory $N \to \infty$. The plot is made for $\epsilon = 0.3$, $N = 40$.}
\end{figure}
\vskip 5mm

Before closing we remark on the limit  $d\to 3$, in which case the Gross-Neveu model describes a real physical system in three dimensions.
The pole contribution~\eqref{quark-singular} vanishes at $d = 3$.
Therefore, the anomalous dimensions $\gamma_q(s), \gamma_\sigma(s)$ and the mass correction $\delta m_q^2(s)$ in Eq.~\eqref{GNm2q} are analytic functions at $s=0$.
Thus, if we do not take into account higher orders in $1/N$ expansion, the anomalous dimensions $\gamma_q(s)$, and $\gamma_\sigma(s)$ cross each other at $s=0$ without branching. 

Second, we have observed that the absence of a  pole at  $s=2-\mu$ in $\delta m^2_q(s)$ does not depend on the signature 
and the flavor structure of the operators.  In the Gross-Neveu model the anomalous dimensions of the negative signature operators have
additional poles at $s=0$, so that $\delta m^2_q(s)$ is singular at this point. In the $\epsilon$ expansion these poles merge so that
$\delta m_q^2(s)$ has a pole at $s=0$ for the negative signature operators.
The situation is different for the $O(N)$-symmetric $\varphi^4$ model.
It can be checked that in this model the anomalous dimensions do not have pole at $s=0$ in the $1/N$ expansion~\cite{Manashov:2017xtt}. 
Therefore $\delta m^2(s)$ for all types of the twist-two operators are regular, both at $s=2-\mu$ and at $s=0$, 
and hence should be regular at $s=0$ in the $\epsilon$ expansion. 
That is indeed  the case as we  saw in Sect.~\ref{sect:phi4}.

\section{Summary}\label{sect:summary}
The anomalous dimensions of the leading twist operators become, as a rule, singular at small positive or negative values of spin.
These singular contributions can be resummed with a variety of techniques~\cite{Velizhanin:2011pb,Vogt:2012gb,Velizhanin:2014dia,Davies:2022ofz}.
In the approach via the generalized double-logarithmic equation the resummed anomalous dimensions have square root singularities in the vicinity of these points~\cite{Velizhanin:2011pb,Velizhanin:2014dia}.
It has been argued in Ref.~\cite{Caron-Huot:2022eqs} that such a structure naturally arises in CFT at the intersection points of different trajectories
which are the scaling dimensions of the operators as function of spins.
Technically, it results in a prediction that a certain quadratic combination of the scaling (anomalous) dimensions of operators is free from singularities in all orders in the perturbation theory.
In particular, it means that the poles at $s=0$ in the anomalous dimensions at high orders are entirely determined from the data at lower orders.

We have studied the behavior of the anomalous dimensions of leading twist operators in several models, the $O(N)$-symmetric $\varphi^4$ model
and the $\varphi^3$ model in the $\epsilon$ expansion near $d=4$ and $d=6$, respectively, and the GNY model in the
$\epsilon$ and $1/N$ expansions.
We have found that in the $\varphi^4$ model the quadratic combination of the anomalous dimensions~\eqref{deltam2} remains finite for all type of twist-two operators in the four-loop approximation.
In the $\varphi^3$ model and GNY model this holds only for the positive
signature anomalous dimensions, while the mass corrections in Eq.~\eqref{deltam2}  are singular at $s=0$ for the negative
signature operators, a finding that agrees with the results of Refs.~\cite{Kirschner:1982qf,Kirschner:1982xw,Kirschner:1983di}.
In the $1/N$ expansion, instead of a single singularity at $s=0$, one finds
three singular points at $s = \epsilon$, $s=0$ and $s=-\epsilon$.
In the limit $\epsilon \to 0$, these singularities merge together within the $\epsilon$ expansion.
In the language of Ref.~\cite{Caron-Huot:2022eqs} each of these points corresponds to the
intersection of the Regge trajectories of certain operators.
We have constructed the corresponding trajectories in the vicinity of the singular points, and have shown that the anomalous dimension of the local operator 
$\sigma^2$ lies on the trajectory of the operator $\sigma \partial_{\mu_1}\ldots\partial_{\mu_s} \sigma$.

In the gauge theories, in QCD and $N=4$ SYM, the mass corrections~\eqref{deltam2} remain finite at the two- and three-loop level, respectively.
Beginning at three and four loops, poles cancel only in the planar sector.
Understanding the underlying reasons for this phenomenon would be highly interesting.
Gaining such insights could significantly aid ongoing efforts to compute the four-loop anomalous dimensions of twist-two operators in QCD.

\appendix

\section*{Acknowledgments}

We are grateful to Vitaly Velizhanin for bringing this topic to our attention.
We would also like to thank him and Vladimir Braun, Sergey Derkachov, Gregory Korchemsky and Petr Kravchuk for numerous and illuminating discussions. 
This work was supported by Deutsche Forschungsgemeinschaft (DFG) through the
Research Unit FOR 2926, ``Next Generation pQCD for Hadron Structure: Preparing for the EIC'', project number 40824754, 
and the ERC Advanced Grant 101095857 {\it Conformal-EIC}. 

\appendix
\addcontentsline{toc}{section}{Appendices} 
\section{Anomalous dimensions in the $O(N)$-symmetric $\varphi^3$ model\label{app:thre-loop-res}}
We present the three-loop results for the anomalous dimensions of the operators considered in the Section~\ref{phi3-model-ii}. For the
operators $\overline{\mathcal{O}}_s^{(ab)}$ and $\overline{\mathcal{O}}_s^{[ab]}$  we have obtained
\begin{align}
    \bar{\gamma}^{(ab)}(s) &= 2\gamma_{\varphi} + \dfrac{2u (-1)^s}{s(s + 1)}
    + \dfrac{2u^2}{s^2(s + 1)^2}\left(1 - \dfrac{1 + 2s}{s(s + 1)}\right)     \notag \\
    &\quad  + \dfrac{2u^2(-1)^s}{3s(s + 1)}\Biggl(S_1
    + \dfrac{2(N + 2)s + N}{4s(s + 1)} -\dfrac{2N + 8}{3}\Biggr)    \notag \\
    &\quad
    + \dfrac{u^3}{s^2(s + 1)^2}\Biggl\{12S_3 + 24S_{1,-2} - 12S_{-3} - \dfrac{2}{3}S_2
    - 12\left(\dfrac{1}{s(s + 1)}  +2\right)S_{-2}
    - 36\zeta_3     \notag \\
    &\quad + \dfrac{4}{3}\left(10 - \dfrac{1 + 2s}{s(s + 1)}\right)S_1
     - \dfrac{24\left(S_{-2} + 1\right)}{(s -1)(s + 2)} + \dfrac{4s - N}{2s^2(s + 1)^2} + \dfrac{(22N + 76)s + 2 - 4N}{9s(s + 1)}     \notag \\
    &\quad
    - \dfrac{2(4N + 13)}{9}\Biggr\}     \notag \\
    &\quad +\dfrac{u^3 (-1)^s}{s(s + 1)}\Biggl\{\dfrac{N + 2}{18}S_2 + \dfrac{1}{9}S_1^2 - 12\left(
    \dfrac{1}{s^2(s + 1)^2} + \dfrac{1}{s(s + 1)}\right)S_{-2} - 4\zeta_3     \notag \\
    &\quad + \left(\dfrac{N + (2N + 4)s}{18s(s + 1)} - \dfrac{302 + 27N}{108}\right)S_1 + \dfrac{12(S_{-2} + 1)}{(s - 1)(s + 2)}
    + \dfrac{4}{s^4(s + 1)^4}    \notag \\
    &\quad + \dfrac{N^2 - 8(N + 2)s + 288}{72s^2(s + 1)^2} + \dfrac{N^2(1 - 16s)
    - 18N(1 + 9s) - 4(259s + 744)}{216s(s + 1)}     \notag \\
    &\quad + \dfrac{8 - 12s}{s^3(s + 1)^3} + \dfrac{12N^2 + 631N + 7099}{648}\Biggr\} + O(u^4),
\end{align}
\begin{align}
    \bar{\gamma}^{[ab]}(s) &= 2\gamma_{\varphi} - \dfrac{2u (-1)^s}{s(s + 1)} + \dfrac{2u^2}{s^2(s
    + 1)^2}\left(1 - \dfrac{1 + 2s}{s(s + 1)}\right)   \notag \\
    &\quad - \dfrac{2u^2 (-1)^s}{3s(s + 1)}\Biggl(S_1
    + \dfrac{2(N + 2)s + N}{4s(s + 1)} - \dfrac{2N + 8}{3}\Biggr)   \notag \\
    &\quad
    + \dfrac{u^3}{s^2(s + 1)^2}\Biggl\{4S_3 + 8S_{1,-2} - 4S_{-3} - \dfrac{2}{3}S_2
    - 4\left(\dfrac{1}{s(s + 1)} + 2\right)S_{-2} \notag \\
    &\quad
    - 12\zeta_3    + \dfrac{4}{3}\left(4 - \dfrac{1 + 2s}{s(s + 1)}\right)S_1
    - \dfrac{8\left(S_{-2} + 1\right)}{(s -1)(s + 2)} + \dfrac{4s - N}{2s^2(s + 1)^2}   \notag \\
    &\quad + \dfrac{(22N + 76)s + 2 - 4N}{9s(s + 1)} - \dfrac{2(4N + 13)}{9}\Biggr\}
    \notag \\
    &\quad
    -\dfrac{u^3 (-1)^s}{s(s + 1)}\Biggl\{\dfrac{N + 2}{18}S_2
    + \dfrac{1}{9}S_1^2 + 4\left(\dfrac{1}{s^2(s + 1)^2} + \dfrac{1}{s(s + 1)}\right)S_{-2}
    \notag \\
    &\quad
     - 4\zeta_3 +\left(\dfrac{N + (2N + 4)s}{18s(s + 1)} - \dfrac{302 + 27N}{108}\right)S_1
     - \dfrac{4\left(S_{-2} + 1\right)}{(s - 1)(s + 2)} + \dfrac{4}{s^4(s + 1)^4}
     \notag \\
    &\quad + \dfrac{N^2 - 8(N + 2)s + 288}{72s^2(s + 1)^2}
     + \dfrac{N^2(1 - 16s) - 18N(1 + 9s) - 4(259s -120)}{216s(s + 1)}
     \notag \\
    &\quad
    + \dfrac{8 - 12s}{s^3(s + 1)^3}  +\dfrac{12N^2 + 631N + 7099}{648}\Biggr\} + O(u^4).
\end{align}
We have checked that the anomalous dimension of the conserved $O(N)$ current vanishes, $\bar{\gamma}^{[ab]}(s = 2) = 0$.
For the operator $\overline{\mathcal{O}}_s^a$ (see Eq.~\eqref{phi3-nonsinglet-1}) we obtained
 \begin{align}
    \bar{\gamma}^a(s) &= \gamma_{\varphi} + \gamma_{\sigma} + (-1)^s\dfrac{2u}{s(s + 1)}
    + \dfrac{2u^2}{s^2(s + 1)^2}\left(1 - \dfrac{1 + 2s}{s(s + 1)}\right)     \notag \\
    &\quad + (-1)^s\dfrac{u^2}{6s(s + 1)}\Biggl((N + 2)S_1
    +\dfrac{2 + s(6 + N)}{s(s + 1)} - \dfrac{8(N + 4)}{3}\Biggr)     \notag \\
    &\quad + \dfrac{u^3}{s^2(s + 1)^2}\Biggl\{\left(N + 1\right)\Bigg(4S_3 - 4S_{-3}
    + 8S_{1, -2} - 12\zeta_3
    - 4\left(\dfrac{1}{s(s + 1)} + 2\right)S_{-2}     \notag \\
    &\quad
    - \dfrac{8\left(S_{-2} + 1\right)}{(s  - 1)(s + 2)}\Bigg) - \dfrac{1}{3}\Biggl(\dfrac{N + (4 + 2N)s + 2}{s(s + 1)}
    - \left(13N + 14\right)\Biggr)S_1  - \dfrac{N + 2}{6}S_{2}     \notag \\
    &\quad + \dfrac{(N+ 2)s - 2}{2s^2(s + 1)^2}
    + \dfrac{N + 2s(19N + 82) - 14}{18s(s + 1)}  - \dfrac{13N + 58}{18}\Biggr\}
    \notag \\
    &\quad
    +(-1)^s\dfrac{u^3}{s(s + 1)}\Biggl\{\dfrac{(N + 2)(N + 6)}{144}S_2
    + \dfrac{(N + 2)^2}{144}S_1^2 - 4(N + 1)\Biggl(
    \dfrac{1}{s^2(s + 1)^2}     \notag \\
    &\quad  + \dfrac{1}{s(s + 1)}\Biggr)S_{-2} - 4\zeta_3 + \left((N + 2)\dfrac{(N + 6)s + 2}{72s(s + 1)}
    - \dfrac{8N^2 + 81N + 518}{216}\right)S_1
    \notag \\
    &\quad
    + \dfrac{4(N + 1)\left(S_{-2} + 1\right)}{(s- 1)(s + 2)} +\dfrac{4}{s^4(s + 1)^4} + \dfrac{4(2 - 3s)}{s^3(s + 1)^3}
    - \dfrac{(N^2 + 8N + 12)s - 292}{72s^2(s + 1)^2}
    \notag \\
    &\quad
    - \dfrac{(8N^2 + 135N + 1122)s - 3N^2 + 861N + 1298}{216s(s + 1)} + \dfrac{12N^2 + 631N + 7099}{648}\Biggr\} + O(u^4).
\end{align}
For  $N = 2$ we have checked that $\bar{\gamma}^a(s) = \bar{\gamma}^{(ab)}(s)$ as it is required by the symmetry.

For the singlet operators $\overline{\mathcal{O}}_s$ and $\Sigma_s$  the anomalous dimensions matrix takes the form
\begin{equation}
    \widehat{\gamma}(s) = \begin{pmatrix}
        \gamma_{\varphi\varphi}(s) && \gamma_{\varphi\sigma}(s) \\
        \gamma_{\sigma\varphi}(s) && \gamma_{\sigma\sigma}(s)
    \end{pmatrix},
\end{equation}
where
\begin{align}
    \gamma_{\varphi\varphi}(s) &= 2\gamma_{\varphi} + (-1)^s\dfrac{2u}{s(s+ 1)}
    + \dfrac{2(N + 1)u^2}{s^2(s + 1)^2}\left(1 - \dfrac{1 + 2s}{s(s + 1)}\right)     \notag \\
    &\quad + (-1)^s\dfrac{2u^2}{3s(s + 1)}\Biggl(S_1 -\dfrac{2N + 8}{3}
    + \dfrac{2(N + 2)s + N}{4s(s + 1)} \Biggr)     \notag \\
    &\quad + \dfrac{u^3}{s^2(s + 1)^2}\Biggl\{12S_3
    + 24S_{1,-2} - 12S_{-3} - \dfrac{(N + 2)(N + 4)}{12}S_2      \notag \\
    &\quad
    - 12\Biggl(\dfrac{1}{s(s + 1)}+2\Biggr)S_{-2} - 36\zeta_3 - \dfrac{N(N - 2)}{12}S_1^2
    + \Biggl(\dfrac{7N^2 - 2N + 120}{9}     \notag \\
    &\quad
    - \dfrac{(5N^2 + 6N + 16)s + 2(N^2 + 2N + 4)}{6s(s + 1)}\Biggr)S_1
    - \dfrac{24\left(S_{-2} + 1\right)}{(s -1)(s + 2)}     \notag \\
    &\quad + \dfrac{3(N^2 + 2N + 4)s - N(N + 7)}{6s^2(s + 1)^2}
    + \dfrac{4(N + 2)(13N + 19)s - 5N^2 - 10N + 4}{18s(s + 1)}
     -     \notag \\
    &\quad \dfrac{2(25N^2 + 25N + 39)}{27}\Biggr\}
     \notag \\
    &\quad
    +(-1)^s\dfrac{u^3}{s(s + 1)}\Biggl\{\dfrac{N + 2}{18}S_2 + \dfrac{1}{9}S_1^2
    - 12\left(\dfrac{1}{s^2(s + 1)^2} + \dfrac{1}{s(s + 1)}\right)S_{-2}
    \notag \\
    &\quad
    - 4\zeta_3 +\left(\dfrac{N + (2N + 4)s}{18s(s + 1)}
    - \dfrac{302 + 27N}{108}\right)S_1 + { \dfrac{12(S_{-2} + 1)}{(s - 1)(s + 2)}} + \dfrac{4(2N + 1)}{s^4(s + 1)^4}
    \notag \\
    &\quad
    -4\dfrac{(2N + 1)(3s - 2)}{s^3(s + 1)^3} + \dfrac{N^2 + 576N + 288 - 8(N + 2)s}{72s^2(s + 1)^2}
\notag\\
&\quad
    + \dfrac{N^2(1 - 16s)
    - 18N(1 + 9s) - 4(259s + 744)}{216s(s + 1)}
    + \dfrac{12N^2 + 631N + 7099}{648}\Biggr\} + O(u^4),
\end{align}
\begin{align}
    \gamma_{\varphi\sigma}(s) &=
     (-1)^s\dfrac{2uN}{s(s + 1)} + \dfrac{2u^2N}{s^2(s + 1)^2}\left(1 - \dfrac{1 + 2s}{s(s + 1)}\right)
          \notag \\
      &\quad
     + (-1)^s\dfrac{2u^2N}{3s(s + 1)}\Biggl(S_1 + \dfrac{1 + 4s}{2s(s + 1)} - 4\Biggr)
           \notag \\
      &\quad
      + \dfrac{u^3N}{s^2(s + 1)^2}\Biggl\{4S_3 + 8S_{1, -2} - 4S_{-3}
      - \dfrac{2}{3}S_2  - 4\left(\dfrac{1}{s(s + 1)}  +2\right)S_{-2}
      -12\zeta_3       \notag \\
      &\quad + \dfrac{4}{3}\left(4- \dfrac{1 + 2s}{s(s + 1)}\right)S_1 - \dfrac{8\left(S_{-2} + 1\right)}{(s - 1)(s + 2)}
      +\dfrac{12s - N - 4}{6s^2(s + 1)^2} + \frac{N (22 s-7)+436 s-10}{36 s (s+1)}
      \notag \\
      &\quad
      +\dfrac{N - 128}{27}\Biggr\}
      \notag \\
      &\quad
      + (-1)^s\dfrac{u^3N}{s(s + 1)}\Biggl\{\dfrac{N + 14}{72}S_2 + \dfrac{N + 6}{72}S_1^2
      - 4\left(\dfrac{1}{s^2(s + 1)^2} + \dfrac{1}{s(s + 1)}\right)S_{-2}
       \notag \\
       &\quad
       - 4\zeta_3 + \left(\dfrac{(N + 14)s + 4}{36s(s + 1)}
       - \dfrac{19N + 318}{108}\right)S_1
       + \dfrac{4\left(S_{-2} + 1\right)}{(s - 1)(s + 2)} + \dfrac{4 (N+1)}{s^4 (s+1)^4}
       \notag \\
       &\quad
       - \dfrac{4 (N+1) (3 s-2)}{s^3 (s+1)^3} +\dfrac{N (289-2 s)-28 s+290}{72 s^2 (s+1)^2}
       - \dfrac{(19N+318) s + N + 318}{54s(s + 1)}
       \notag \\
       &\quad
       + \frac{583 N+10046}{864}\Biggr\} + O(u^4),
\end{align}
\begin{align}
    \gamma_{\sigma\varphi}(s) &=
     (-1)^s\dfrac{2u}{s(s + 1)} + \dfrac{2u^2}{s^2(s + 1)^2}\left(1 - \dfrac{1 + 2s}{s(s + 1)}\right)       \notag \\
      &\quad
     + (-1)^s\dfrac{u^2}{3s(s + 1)}\Biggl(NS_1 + \dfrac{(N + 2)s + 1}{s(s + 1)}
      - \dfrac{8(N + 1)}{3}\Biggr)       \notag \\
      &\quad  + \dfrac{u^3}{s^2(s + 1)^2}\Biggl\{4S_3 + 8S_{1, -2} - 4S_{-3}
       + \dfrac{(N - 2)}{12}S_1^2 - \dfrac{3N + 2}{12}S_2  - 12\zeta_3
       \notag \\
       &\quad
       - 4\left(\dfrac{1}{s(s + 1)}  +2\right)S_{-2}
       -\left(\dfrac{N (3 s+2)+10 s+4}{6 s (s+1)} + \dfrac{N - 50}{9}\right)S_1
       \notag \\
       &\quad
       - \dfrac{8\left(S_{-2} + 1\right)}{(s - 1)(s + 2)} + \dfrac{(3s - 1)N + 6s - 4}{6s^2(s + 1)^2}
       +\frac{N (114 s+5)+252 s-34}{36 s (s+1)} - \dfrac{14(N + 7)}{27}\Biggr\}
       \notag
       \\&\quad
    +(-1)^s\dfrac{u^3}{s(s + 1)}\Biggl\{\dfrac{N(3N + 10)}{144}S_2
    + \dfrac{N(3N + 2)}{144}S_1^2 - 4\Biggl(\dfrac{1}{s^2(s + 1)^2}
     \notag \\
    &\quad+ \dfrac{1}{s(s + 1)}\Biggr)S_{-2}  - 4\zeta_3 + \left(\dfrac{N ((3 N+10) s+4)}{72 s (s+1)}
    - \dfrac{6N^2 + 23N + 108}{54}\right)S_1  \notag \\
    &\quad+ \dfrac{4\left(S_{-2} + 1\right)}{(s - 1)(s + 2)}
    +\frac{4 (N+1)}{s^4 (s+1)^4} -\frac{4 (N+1) (3 s-2)}{s^3 (s+1)^3}
    -\frac{3 N^2 s-N (287-10 s)-294}{72 s^2 (s+1)^2}
    \notag \\
    &\quad
    +\frac{N^2 (9-24 s)-2 N (81 s+1)-4 (251 s+328)}{216 s (s+1)} \notag \\
    &\quad  + \frac{144 N^2+3107 N+26846}{2592}\Biggr\} + O(u^4),
\end{align}
\begin{align}
    \gamma_{\sigma\sigma}(s) &=
    2\gamma_{\sigma} + \dfrac{2u^2N}{s^2(s + 1)^2}\left(1 - \dfrac{1 + 2s}{s(s + 1)}\right)
    + \dfrac{4u^3N}{s^2(s + 1)^2}\Biggl\{S_{3} + 2S_{1, -2} - S_{-3} - \dfrac{3N + 2}{48}S_2
    \notag \\
    &\quad
    - \left(\dfrac{1}{s(s + 1)} + 2\right)S_{-2} - 3\zeta_3
    + \dfrac{N  - 2}{48}S_1^2 - \Biggl(\dfrac{(3N + 10)s + 2N + 4}{24s(s + 1)} + \dfrac{N - 50}{36}\Biggr)S_1
    \notag \\
   &\quad - \dfrac{2\left(S_{-2} + 1\right)}{(s - 1)(s + 2)}  + \dfrac{3s(N + 2) + N - 8}{24s^2(s + 1)^2}
   +\dfrac{24(N + 8)s + 7N - 26}{72s(s + 1)} +\dfrac{11N - 148}{108}\Biggr\}
   \notag \\
   &\quad + (-1)^s\dfrac{4u^3N}{s^3(s + 1)^3}\Biggl\{S_{-2} + \dfrac{4\left(S_{-2}
    + 1\right)}{(s - 1)(s + 2)}+ \dfrac{1}{s^2(s + 1)^2} + \dfrac{2 - 3s}{s(s + 1)} + 3\Biggl\}.
\end{align}
We have checked that the anomalous dimensions of the conserved $U(1)$ current and the stress-energy tensor vanish,
i.e., projecting with the vectors $v_1^T=(1,-2)$ and $v_2^T=(1,1)$,
\begin{align}
  v_1^T \cdot \widehat{\gamma}(s = 2) &= 
    \begin{pmatrix}
        1, && - 2
    \end{pmatrix} \cdot \widehat{\gamma}(s = 2) = 0,
\nonumber \\
v_2^T \cdot \widehat{\gamma}(s = 3) &=
  \begin{pmatrix}
        1, && 1
    \end{pmatrix} \cdot \widehat{\gamma}(s = 3) = 0.
\end{align}
%

\renewcommand{\theequation}{\Alph{section}.\arabic{equation}}
\renewcommand{\thetable}{\Alph{table}}
\setcounter{section}{0} \setcounter{table}{0}

\newpage
\bibliography{refs.bib}
\bibliographystyle{JHEP}

\end{document}